\documentclass[11pt]{aastex}
\usepackage{graphicx,emulateapj5,apjfonts}
\usepackage{onecolfloat}
\usepackage{epsf}

\shortauthors{Bell et al.}
\shorttitle{Galaxy Luminosities and Stellar Masses}

\newcommand{\nd}{\nodata}

\newcommand{\hi}{{\rm H{\sc i} }}

\newcommand{\pegase}{{\sc P\'egase }}

\slugcomment{{\sc Astrophysical Journal Supplement Series, in press: } August 4, 2003 }

\begin{document}


\def\head{

\title{The Optical and Near-Infrared Properties of Galaxies: I. 
Luminosity and Stellar Mass Functions }

\author{Eric F.\ Bell}
\affil{Max-Planck-Institut f\"ur Astronomie,
K\"onigstuhl 17, D-69117 Heidelberg, Germany; \texttt{bell@mpia.de}}
\author{Daniel H.\ McIntosh, Neal Katz, and Martin D.\ Weinberg}
\affil{Department of Astronomy, University of Massachusetts, 
710 North Pleasant Street,
Amherst, MA 01003-9305; \texttt{dmac@hamerkop.astro.umass.edu, 
nsk@kaka.astro.umass.edu, weinberg@astro.umass.edu}}

\begin{abstract}
We use a large sample of galaxies from the {\it Two Micron All
Sky Survey} (2MASS) and the {\it Sloan Digital Sky Survey} (SDSS)
to calculate galaxy 
luminosity and stellar mass functions in the local Universe.
We estimate
corrections for passband shifting and galaxy evolution, 
as well as present-day stellar mass-to-light (M/L) ratios,
by fitting the optical--near-infrared
galaxy data with simple models.  
Accounting for the 8\% galaxy overdensity in the 
SDSS early data release region, the optical and 
near-infrared luminosity functions we construct for this sample
agree with most recent literature
optical and near-infrared determinations within the uncertainties.  
We argue that 
2MASS is biased against low surface brightness galaxies, 
and use SDSS plus our knowledge of stellar populations to 
estimate the `true' $K$-band luminosity function.  This 
has a steeper faint end slope and a slightly higher 
overall luminosity density than the direct estimate.
Furthermore, assuming a universally-applicable stellar initial 
mass function (IMF), we find good agreement
between the stellar mass function we derive from the 2MASS/SDSS data
and that derived by Cole et al. (2001; MNRAS, 326, 255).  
The faint end slope slope for the stellar mass function
is steeper than $-1.1$, reflecting the low stellar M/L ratios characteristic
of low-mass galaxies.  
We estimate an upper limit to the stellar mass density in the local Universe  
$\Omega_* h = 2.0\pm0.6\times 10^{-3}$ by assuming an IMF 
as rich in low-mass stars as allowed by observations of galaxy dynamics 
in the local Universe.  The stellar mass
density may be lower than this value if a different IMF
with fewer low-mass stars is assumed. 
Finally, we examine type-dependence in the optical and near-infrared
luminosity functions and the stellar mass function.
In agreement with previous work, we
find that the characteristic luminosity or mass of early-type
galaxies is larger than for later types, and the faint end 
slope is steeper for later types than for earlier types.
Accounting for typing uncertainties, we estimate
that at least half, and perhaps as much as 3/4, 
of the stellar mass in the Universe is in 
early-type galaxies.

As an aid to workers in the field, we present in an appendix
the relationship
between model stellar M/L ratios and colors in SDSS/2MASS passbands,
an updated discussion of near-infrared stellar M/L ratio estimates,
and the volume-corrected distribution of $g$ and $K$-band
stellar M/L ratios as a function of stellar mass.
  
\end{abstract}

\keywords{galaxies: luminosity function, mass function -- galaxies: general 
--- galaxies: evolution --- galaxies: stellar content}
}

\twocolumn[\head]

\section{Introduction}

The distribution of galaxy luminosities and stellar
masses in the present-day Universe is 
of fundamental importance for studying 
the assembly of galaxies over cosmic time,
both observationally and theoretically
\citep[e.g.,][]{lilly95,lin99,cole00,brinchmann00,somerville01,wolf03}.  
In addition to providing the 
zero redshift baseline for luminosity function (LF) evolution,
the local LF constrains powerfully 
much of the important physics affecting the assembly of 
baryons in dark matter halos.  For example, gas accretion
and cooling dominates the bright end of the LF, whereas
feedback and photoionization affect primarily 
fainter galaxies \citep[e.g.,][]{cole00,benson02}.  Near-infrared (NIR)
luminosities of galaxies are
particularly useful as the mass-to-light (M/L) ratios in 
the NIR vary only by a factor of two or less across a wide range
of star formation (SF) histories \citep[see also the Appendix]{ml}, 
contrasting with a factor of 
ten change in M/L ratio at the blue end of the optical regime.  
Therefore, NIR luminosities provide a 
cleaner estimate of galaxy stellar masses, which are 
more robustly predicted by the
theoretical models \citep[e.g.,][]{gardner97,cole01,kochanek01}.
The goal of this paper is to use the NIR {\it Two Micron All Sky Survey}
\citep[2MASS;][]{skrut}
in conjunction with optical data and redshifts from the {\it Sloan
Digital Sky Survey} \citep[SDSS;][]{york00} to explore
the distribution of galaxy luminosities in the optical and 
NIR, and to use these data to estimate the distribution of
stellar masses in the local Universe.

There have been a number of recent studies that have 
estimated LFs and mass functions (MFs), 
based on a number of recent large surveys.
Around the knee of the LF, which represents
the dominant contribution to the overall luminosity density, the agreement 
between the LFs from different surveys is good.
In the optical, luminosity densities agree at typically
the 20\% level or better,
accounting for differences in filter bandpasses and median
redshift \citep[e.g.,][]{norberg02,liske03,blanton03}.
A similar conclusion is found for the 
NIR $K$-band \citep[e.g.,][]{gardner97,cole01,kochanek01}. 
There are some indications that the behavior of the difficult-to-measure
fainter galaxies may depend on environment \citep{tully02},
although these galaxies
do not exist in sufficient numbers to 
contribute significantly to the luminosity density of the local
Universe \citep[e.g.,][]{zabludoff00,trentham02}.

Three notable exceptions to this concordance of recent LF measurements
are the {\it Las Campanas Redshift Survey} \citep[LCRS;][]{lin96},
the early SDSS LF from \citet{blanton01}, and the $K$-band LF estimate
of \citet{huang03}.  
The LCRS estimates are relatively consistent with 
more recent estimates of the optical LFs \citep[e.g.,][]{blanton03}, but
because of two offsetting effects: ({\it i}) the neglect of 
evolution, which biases the luminosity density to higher values; and ({\it ii})
the use of isophotal magnitudes, which biases the luminosity density
back down to lower values \citep{blanton01,blanton03}.  \citet{blanton01}
find $\ga$50\% more luminosity density in the local Universe
than more recent SDSS or {\it Two Degree Field Galaxy Redshift Survey}
\citep[2dFGRS;][]{colless01} estimates. 
This offset is due mostly to the neglect of galaxy evolution 
and partially to the use of crude $k$-corrections
\citep{blanton03}.  The difference between the $K$-band
LF of \citet{huang03} and other local estimates is less
well-understood, but could stem from the neglect of evolution 
corrections, LF fitting uncertainties and/or large-scale structure
(their LF estimate comes from an area of sky 50 times smaller than 
the area studied in this work; we discuss this issue in 
more detail in \S \ref{sec:k}).

Furthermore, it is unclear if the 
optical and NIR LFs are mutually consistent. \citet{cole01}
compared the optical $z$-band LF from \citet{blanton01} with 
their hybrid $J/K$-band LF, finding poor agreement.   
\citet{wright01} finds over a factor of two offset between
extrapolations from the optical LFs of \citet{blanton01}
and 2MASS-derived $K$-band LFs \citep{cole01,kochanek01}.
Given the above argument that luminosity densities
in the optical and NIR are basically known to within
20\%, it is unclear 
whether this discrepancy can be simply accounted for
by the neglect of evolution corrections by \citet{blanton01}, or whether,
for example, this is an indication of gross global incompleteness in $K$-band 
LFs.  Furthermore, the landmark stellar MFs derived by \citet{cole01}
have not been, as yet, tested systematically.

In this paper, the first in a series of papers focusing on the 
optical and NIR properties of galaxies in the local Universe, 
we use the NIR 2MASS 
in conjunction with optical data and redshifts from SDSS to 
explore in detail the LFs of 
galaxies over a factor of 6 in wavelength from the $u$-band 
(0.35\micron) to the $K$-band (2.15\micron).  
We then, following the methodology of \citet{bdj,ml}, 
use the constraints on the optical-NIR
spectral energy distributions (SEDs) in conjunction with state-of-the-art
stellar population synthesis (SPS) models to 
investigate in detail the stellar MF of galaxies over
a factor of 1000 in stellar mass, assuming a universally-applicable 
stellar initial mass function (IMF).
We used these stellar mass estimates in 
conjunction with a statistically-estimated
cold gas masses (H{\sc i} and H$_2$) to construct
the cold baryonic MF in the local
Universe and the efficiency of galaxy formation 
\citep[see also, e.g., Salucci \& Persic 1999]{bary}.
In subsequent papers we will examine, e.g., the $K$-band 
size distribution of galaxies, the photometric properties
of a $K$-selected sample, the dust contents
and SF histories of disk galaxies, and the $K$-band LF of 
bulges and disks separately, amongst other goals.

This paper is arranged as follows.  In \S \ref{sec:data}, we discuss
the data, focusing on the most important sources
of error and incompleteness.  
In \S \ref{sec:method}, we discuss our method for 
deriving $k$-corrections, evolution corrections and stellar M/L ratios.
In \S \ref{sec:lf}, we construct and discuss optical and NIR LFs for 
our sample of galaxies.  We construct 
stellar MFs in \S \ref{sec:mf} and discuss these further in \S \ref{sec:disc}.
We summarize in \S \ref{sec:conc}.
In the Appendix, we present the distribution of 
color-derived stellar M/L ratio estimates as a function of 
galaxy mass and fits to the color-M/L ratio correlations
in the SDSS/2MASS passbands as aids to workers in the field.
We assume 
$\Omega_{\rm matter} = 0.3$, $\Omega_{\Lambda} = 0.7$, and 
$H_0 = 100 h $\,km\,s$^{-1}$\,Mpc$^{-1}$.  For estimating 
evolution corrections,
we assume $h = 0.7$.  Sections \ref{sec:data}, \ref{sec:method}, and 
the Appendix go into considerable detail regarding the uncertainties
and stellar M/L ratios; thus, readers interested mainly in the 
results should read \S \ref{sec:over} and then skip 
directly to \S \ref{sec:lf}.

\section{The Data, Data Quality, and Selection Effects} \label{sec:data}

\subsection{Overview} \label{sec:over}

We use the SDSS Early Data Release \citep[EDR;][]{edr}
to provide a nearly complete $13 \le r \le 17.5$ sample of 
22679 galaxies over 414 square degrees with accurate $ugriz$ 
fluxes and magnitudes.  We match these SDSS spectroscopic
sample galaxies with the 2MASS extended source catalog 
\citep[XSC;][]{jarrett00} and point source catalog 
(PSC)\footnote{http://www.ipac.caltech.edu/2mass/releases/second/doc/ancillary/pscformat.html}.
To match the catalogs, 
we choose the closest galaxy within 2\arcsec as the best
match \citep[for reference, 
the random and systematic positional uncertainties 
of 2MASS and SDSS are
$\la$200 and $\sim$50 milliarcseconds, respectively;][]{pier03}.
In this way, we have a reasonably complete $13 \le r \le 17.5$ sample of galaxies
with redshifts, 12085 of which have a match in 
the 2MASS XSC (and therefore have
$ugrizK$ fluxes, half-light radii and concentrations 
in $r$ and $K$-bands), 6629 of which have a match in 
the 2MASS PSC (and therefore have
$ugrizK$ fluxes, and half-light radii and concentration parameters in 
$r$-band), and 3965 of which have no match in either
2MASS catalog (and thus have the optical data only).  
We choose to use only the 2MASS $K$-band at the present
time.

A complete description of these catalogs is far beyond the 
scope of this paper \citep[see e.g.,][for 
more details]{jarrett00,blanton01,cole01,edr}.  Here,
we discuss the most important aspects for our purposes:
the accuracy of the magnitudes, concentrations and surface brightnesses, 
and the completeness of the catalogs.

\subsection{Magnitude Accuracy} \label{magacc}

\begin{figure}[tb]
\vspace{0.1cm}
\hspace{-0.5cm}
\epsfxsize=8.5cm
\epsfbox[120 170 417 467]{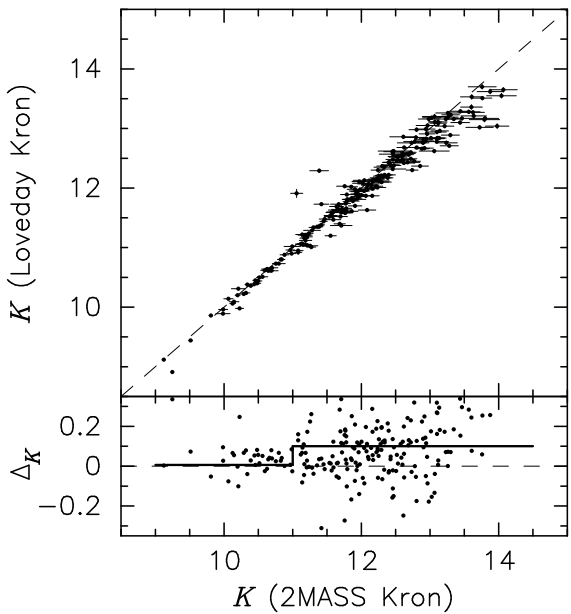}
\vspace{-0.2cm}
\caption{\label{fig:loveday} Comparison of 2MASS $K$-band Kron 
magnitudes with deep $K$-band magnitudes for 
223 galaxies from \citet{loveday}.  The lower panel shows
the difference (2MASS$ - $Loveday) in $K$-band magnitudes.  The bold line
represents the mean magnitude difference $0.01\pm0.04$ ($0.10\pm0.02$)
for $K<11$ ($K\geq11$).  The scatter is $\sim0.2$ mag.}
\end{figure}

\begin{figure}[tb]
\vspace{0.1cm}
\hspace{-0.5cm}
\epsfxsize=8.5cm
\epsfbox[120 170 417 467]{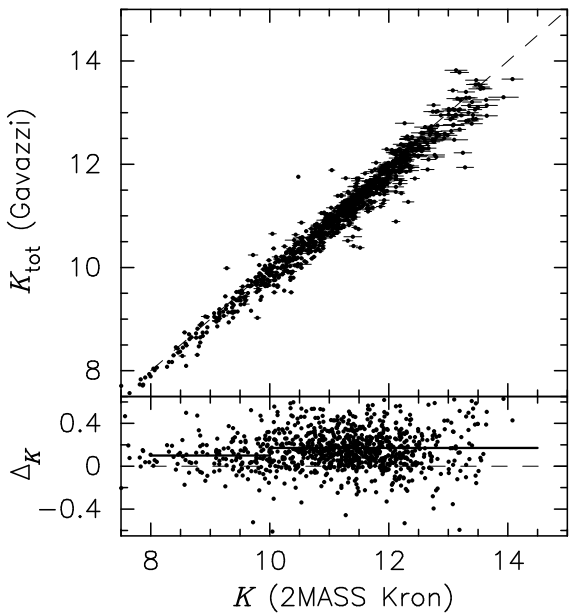}
\vspace{-0.2cm}
\caption{\label{fig:gavazzi} Comparison of 2MASS $K$-band Kron 
magnitudes with deep $K$-band magnitudes (transformed from 
$H$-band using the 2MASS measured $H-K$ color) for 
1017 galaxies from papers by Gavazzi and coworkers.  The lower panel shows
the difference (2MASS$ - $Gavazzi) in $K$-band magnitudes.  The bold line
represents the mean magnitude difference $0.06\pm0.02$ ($0.17\pm0.01$)
for $K<10$ ($K\geq10$).  The scatter is $\sim0.2$ mag.}
\end{figure}

An important focus of this paper is the 
discussion of the offset between optical and 
2MASS $K$-band LFs \citep[e.g.,][]{wright01}.
Because the NIR luminosity density seemed a factor
of two below expectations, one of the principal concerns was 
a large shortfall in
either the magnitudes or numbers of galaxies in 
2MASS.  In this section, we discuss the 
accuracy of the 2MASS $K$-band magnitudes in 
detail, and briefly summarize the expected
accuracy of magnitudes in the other passbands. 

In many respects, $K$-band data from 2MASS is the ideal 
tool for constraining galaxy LFs and the stellar MF.  \footnote{We note that,
strictly speaking, 2MASS adopts a $K_s$-band that peaks at rather shorter
wavelengths than the standard $K$-band, but we we will
call it $K$-band for brevity in this paper.}
$K$-band galaxy luminosities are five to ten times less sensitive
to dust and stellar population effects than optical luminosities,
allowing an accurate census of 
stellar mass in the local Universe \citep[e.g.,][]{ml}.
Furthermore, 2MASS
covers the entire sky homogeneously, with 1\% systematic variations in
zero point \citep{nikolaev00}.
However, in the NIR the sky background is roughly a factor of 100 times 
brighter than the mean surface brightness of luminous galaxies,
and the exposure time of 2MASS is short 
\citep[7.8 seconds with a 1.3-m telescope;][]{skrut}.  Thus,
low surface brightness (LSB) features, such as LSB galaxies or
the outer regions of normal galaxies, may be missed
by 2MASS.

To test how much light 2MASS misses in the LSB outer parts
of galaxies, we compare 2MASS $K$-band magnitudes from 
the XSC with $K$-band
magnitudes from deeper imaging data.  Following
\citet{cole01}, we show Loveday's $K$-band Kron\footnote{\citet{kron}
magnitudes are measured in apertures that are related to the 
galaxy radius (for 2MASS, not less than 5\arcsec).}
magnitude from relatively deep data 
(10 minutes on the {\it Cerro Tololo International Observatory} 
1.5-m telescope) 
against 2MASS $K$-band Kron magnitudes (Fig.\ \ref{fig:loveday}).
At $K < 11$, 2MASS Kron magnitudes seem quite
accurate, with a systematic offset of 0.01$\pm$0.04 mag.
At fainter magnitudes, 2MASS Kron magnitudes underestimate the true 
magnitude by 0.10$\pm$0.02 mag (scatter $\sim$0.2 mag).  \citet{cole01}
found a larger offset between Second Incremental Data Release 2MASS
$K$-band and total magnitudes; since the 
Second Incremental Data Release there have been improvements
to the reduction pipeline that have improved the quality of 
2MASS $K$-band Kron magnitudes.  

We check this offset by comparison with a larger sample 
of galaxies imaged in the $H$-band by \citet{gav96a,gav96b,gav00}
and \citet{bos00}\footnote{This sample was used to test
the circular isophotal magnitudes used by \citet{kochanek01} 
for their $K$-band derived LF.}.
We adopt the 2MASS $H-K$ color to estimate 
the total $K$-band magnitude; the typical value is $H-K \sim 0.25$,
almost independent of galaxy type.  The average offset brighter
(fainter) than $K=10$ mag is 0.06 (0.17) mag, in the sense
that Gavazzi's magnitudes are slightly 
brighter than the 2MASS Kron magnitudes (see Fig.\ \ref{fig:gavazzi}).
We do not adopt this correction in this paper owing to uncertainties 
in transforming $H$-band data into $K$-band.  We do, nevertheless,
choose to adopt an
offset of 0.1 mag for all galaxies (not just galaxies
with $K \ge 11$) to better match Gavazzi's offset. 
We note that magnitudes corrected in this way will be within 0.1 mag of 
total, independent of whether one compares
them to Gavazzi's or Loveday's total magnitudes.  We adopt a
0.1 mag uncertainty in the correction to 
total $K$-band fluxes, added in quadrature
with the 2MASS random magnitude error.  We tested whether the 
correction to total is a function of $K$-band surface brightness, 
as one could imagine that the fraction of light lost may be larger
for lower surface brightness galaxies.  We found no correlation 
between the correction to total flux and $K$-band surface brightness
within the errors, supporting our use of a blanket 0.1 mag offset.

We also choose to match to the 2MASS PSC.
There are very few matches to the comparison samples:
8 from \citet{loveday} and 8 from the sample from Gavazzi and
coworkers.  We find
mean offsets of $-0.9$ mag (0.3 mag RMS) and $-0.8$ mag (0.3 mag
RMS) for the two samples.  We disregard two outliers (with no offset and
a $-3.1$ mag offset)
from the eight of the Gavazzi sample.  We account for the large PSC offsets
by subtracting 0.85 mag from the 2MASS PSC
$K$-band magnitudes, and setting their $K$-band errors to 0.5 mag.
These magnitudes are clearly of very limited use. We use
them primarily to constrain only roughly the $k$-correction, evolution 
correction, and stellar M/L ratio estimates.  In particular, our choice
of $K$-band magnitude limit (extinction-corrected $K$-band 
Kron magnitude of 13.57, with the offset included after galaxy 
selection) includes only 66 galaxies from the 
PSC, or just over 1\% of our $K$-band selected sample. 

Because of its high signal-to-noise, 
SDSS Petrosian\footnote{SDSS Petrosian magnitudes are 
estimated within an aperture that is twice the radius 
at which the local surface brightness is 1/5 of the mean 
surface brightness within that radius \citep{strauss02}.}
$ugriz$ magnitudes are expected to be accurate
to better than 0.05 mag in a random and systematic 
sense \citep{strauss02,blanton03}.  Sloan papers typically make the 
distinction between preliminary magnitudes presented 
by the EDR in the natural Sloan 2.5-m telescope system,
denoted $u^*g^*r^*i^*z^*$, and the `true' Sloan magnitudes $ugriz$.
We denote the EDR Petrosian magnitudes $ugriz$ for brevity.
Petrosian magnitudes of well-resolved early-type
galaxies (with close to $R^{1/4}$ law luminosity profiles) 
underestimate the total flux by $\sim$0.1 mag
because their surface brightness profiles fall off
very slowly at large radii \citep{strauss02,blanton03}.  In this paper, we
crudely correct for this effect by subtracting 0.1 mag 
from the magnitude of any galaxy with 
an $r$-band concentration parameter of $c_r > 2.6$ (defined
in the next section).  
While simplistic, it allows us to estimate the 
total fluxes for early-type galaxies to within 0.05 mag.  We adopt
a magnitude error of 0.05 mag for all galaxies, added in 
quadrature to the (tiny) internal SDSS random magnitude errors.
Note that we apply all magnitude offsets {\it after} galaxy selection.

\subsection{Concentration Parameters and Surface Brightnesses} \label{sec:cnc}

In this paper, we study primarily the overall
luminosities and stellar masses of galaxies, choosing not
to focus on their structural parameters, such as 
concentration parameter or surface
brightness.  Nevertheless, we do use concentration parameter
as a crude discriminant between early and late-type galaxies
and surface brightnesses when examining the completeness of 
the galaxy samples.  

We adopt as our primary morphological classifier
the $r$-band concentration parameter, $c_r = r_{90}/r_{50}$, 
where $r_{90}$ and $r_{50}$ are the circular aperture 
radii within which 90\% and
50\% of the Petrosian flux are contained, respectively.
The concentration parameter has been extensively
used within the SDSS collaboration to
separate between early and late-type galaxies in a rudimentary fashion;
early-type galaxies have higher $c_r$ than later types.  This 
is motivated by the work of \citet{strat} 
and \citet{shim}, who find a scattered but 
reasonable correlation between 
qualitative morphological classifications and $c_r$.
\citet{strat} suggest a 
$c_r \ge 2.6$ selection for early-type galaxies and this
cut has been adopted by \citet{kauffmann03a}. 
We also adopt this criterion,
primarily because it is easily reproducible, facilitating
easy comparison with our results by other workers.
\citet{blanton03c} note the sensitivity of the the 
concentration parameter to seeing; more heavily smoothed
early-type galaxies appear less concentrated than they
would be either if they were observed with better seeing or were closer.
Because of this, the early type definition is conservative;
intrinsically smaller or more distant early types maybe 
misclassified as later types owing to seeing effects.

\begin{figure}[tb]
\vspace{-0.5cm}
\hspace{-0.5cm}
\epsfbox{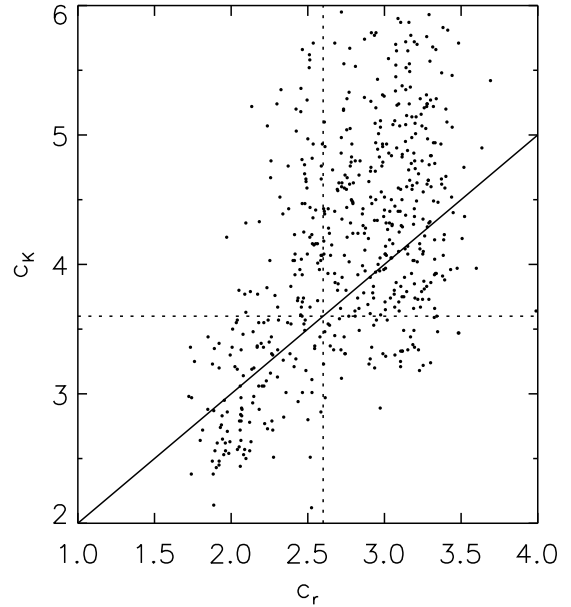}
\vspace{-0.2cm}
\caption{\label{fig:conc} $K$-band concentration
parameter $c_K = r_{75}/r_{25}$ against $r$-band 
concentration parameter $c_r = r_{90}/r_{50}$.  
The solid line denotes $c_K \sim c_r + 1$, and
the dashed lines denote the two rough early-type cuts
at $c_r \ge 2.6$ and $c_K \ge 3.6$. }
\end{figure}

2MASS also gives a concentration parameter $c_K = r_{75}/r_{25}$,
where $r_{75}$ and $r_{25}$ are the elliptical aperture 
within which 75\% and 25\% of the flux are contained.  
In Fig.\ \ref{fig:conc}, we compare the SDSS $c_r$ to the 2MASS
$c_K$ in an effort to explore systematic uncertainties in the 
use of concentration parameters as a morphological typing tool
(we compare $c_r$ to color selection later also in \S \ref{sec:col}).
We restrict our comparison to
$K < 12$ galaxies, which have sufficient S/N to 
estimate $c_K$.
It is clear that there are systematic differences
between the two definitions, which are manifested by
zero point shifts, a non-unity slope, and a substantial scatter.
Nevertheless, making the crude approximation that $c_K \sim c_r + 1$ 
({\it the solid line}), we can compare the fractions classified as early-type
with both definitions.  Using $c_r \ge 2.6$, we find that 401/603 
$K < 12$ EDR galaxies are classified as early-type.
Using $c_K \ge 3.6$, we find that 439/603 galaxies are classified
as early-type.  Furthermore, 355/401 galaxies (89\%$\pm$7\%) of
$r$-band classified early-types are classified as early type
using the $K$-band classification.  

We use $g$, $r$, and $K$-band surface brightnesses only as a rough
check on the completeness properties of the 2MASS-matched
sample and on the $K$-band LF.  These 
surface brightnesses are defined to be the average surface 
brightness within the half-light radii.  Since the magnitudes 
for SDSS and 2MASS XSC are accurate to 
at worst 20\%, we expect that the half-light surface brightnesses
will be accurate to $\la 30$\%, given the 20\% error in 
total magnitude, added in quadrature with the effect of a 20\% scale size
error, which is typical of scale-size comparisons between different authors
\citep[see, e.g.,][]{bdj}.   Accuracies of this order
are more than sufficient for our present purposes.

\subsection{Completeness} \label{samp:compl}

The homogeneity and 
completeness of SDSS and 2MASS make them powerful tools for understanding the
characteristics of galaxies in the local Universe.
To construct meaningful LFs from these datasets, we must 
understand the completeness characteristics of each survey.
We choose SDSS EDR spectroscopic sample galaxies
with Galactic foreground extinction-corrected $13 \le r \le 17.5$
\citep{sfd}, following \citet{edr}.  
This galaxy sample is nearly complete, 
as discussed in much more detail by e.g. \citet{blanton01},
\citet{edr} or \citet{strauss02}.  We find that the area
covered by the SDSS spectroscopic sample is 414 square degrees,
90\% of the 462 square degrees covered by the EDR imaging data
\citep{edr}.  This is quite consistent with the the statement by 
\citet{edr} that only 93\% of the spectroscopic 
tiles were attempted; we adopt the 3\% difference as our systematic error
in determining the sky coverage of this sample.
We estimate a total
completeness within this area by querying SDSS EDR photometric catalog galaxies
satisfying the spectroscopic catalog inclusion criteria
as outlined by \citet{strauss02}.  This value is 78\%, which
is consistent with a 2\% loss of galaxies due to bright stars,
a $> 99$\% redshift success rate, and between 80\% and 90\% 
targeting efficiency \citep[e.g.,][]{blanton01,edr}.
A value of 85\% was recently found by \citet{nakamura03}
for bright SDSS galaxies ($r \le 15.9$) in the EDR: we adopt the 
difference between our and Nakamura et al.'s measurements as
the systematic error in the completeness, which is propagated 
through into the $\phi^*$ and $j$ estimates later.
We do not take account 
of the detailed, position-dependent completeness of the sample.
While a detailed accounting for the completeness as a function of
position is pivotal for estimating galaxy clustering
properties, it is of only minor importance for estimating 
the overall LF.  Finally, we note that there is little systematic bias
within SDSS against galaxies within the selection limits.
Of order 0.1\% of the lowest surface brightness galaxies
are not targeted because a spectrum would be impossible to 
obtain, and because over 3/4 of the lowest surface brightness 
features in the SDSS imaging survey are artifacts \citep{strauss02}.
Also, $\la 5$\% of bright galaxies 
are rejected because they overlap a bright, saturated star, or 
because they have a very bright fiber magnitude and are not targeted
to avoid severe cross-talk between the fiber spectra.  Neither of 
these biases will significantly affect our analysis.

\begin{figure*}[tb]
\center{\includegraphics[scale=1.0, angle=0]{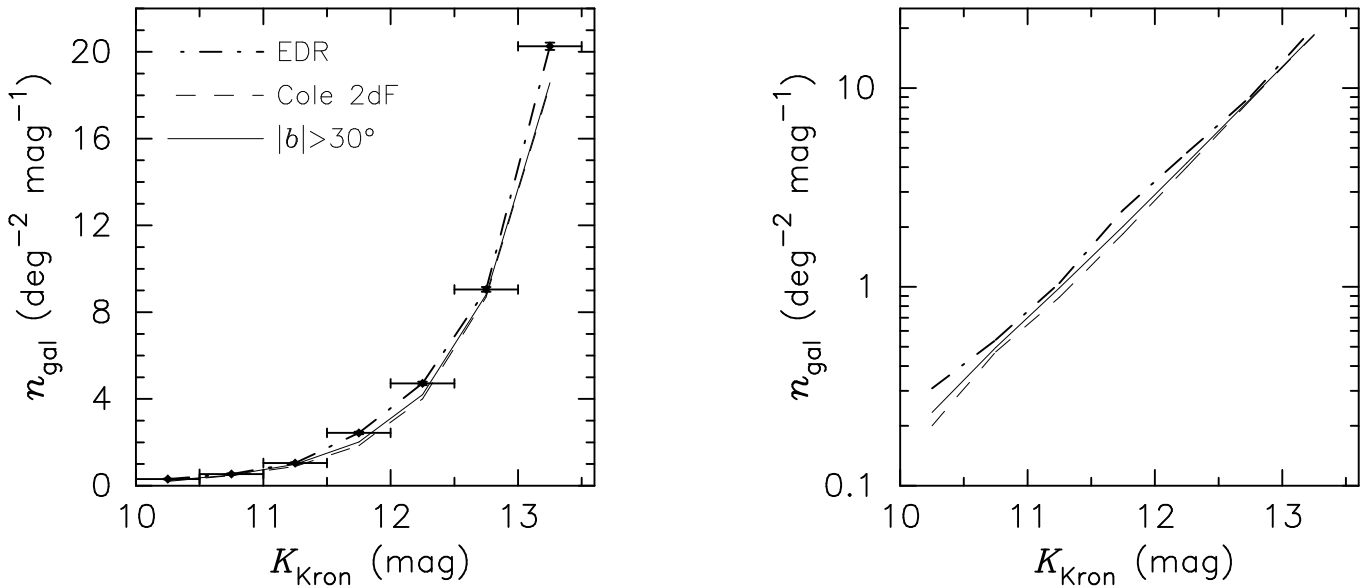}}
\caption{\label{fig:ndensity} Number 
counts of galaxies per square degree per magnitude
as a function of 2MASS $K$-band Kron apparent magnitude for the whole
sky with $|b| \ge 30$ ({\it solid line}), the SDSS EDR ({\it dot-dashed})
and the 2dFGRS ({\it dashed line)}.  Linear ({\it left}) 
and logarithmic ({\it right}) scales are shown. The points
with error bars show the error in the EDR galaxy number density
({\it vertical error bars}) and the magnitude range used for 
each bin ({\it horizontal error bars}).  }
\end{figure*}

In addition to estimating the completeness of SDSS internally,
we determine whether the 
SDSS EDR area is overdense using the full coverage of 2MASS.  We 
estimate overdensities
by comparing the number of 2MASS extended sources with
$10 < K < 13.5$ in the sky outside of the Galactic Plane ($|b| \ge 30\arcdeg$)
with the number of similar sources in the SDSS EDR region.
We use
an area that is slightly less than the 414 square degrees that we calculate
for the spectroscopic 
EDR coverage because we choose
rectangular areas that are fully enclosed by the SDSS EDR boundaries.
We show in Fig.\ \ref{fig:ndensity} that
the EDR is overdense over the entire magnitude
range $10 < K < 13.5$.  We include the estimated density in 
the 2dFGRS region for comparison and give the number counts for each region
in Table \ref{tab:ndensity}.
The SDSS EDR is 8\% overdense (with a 1\% Poisson uncertainty), 
and the 2dFGRS region \citep[used by][]{cole01}
is 2\% underdense, compared to the whole sky.  Although this estimate
is admittedly rough because $10 < K < 13.5$ galaxies 
are a somewhat different set of galaxies than those with $13 \le r \le 17.5$, 
the overdensity estimate is accurate given that we compare to half
of the entire sky ($|b| \ge 30\arcdeg$).  Furthermore, our estimate
is insensitive to Galactic foreground extinction.
We account for the EDR region overdensity
in our analysis by multiplying the effective survey
area by 1.08 when constructing our LFs.

\begin{deluxetable}{lccc}
\tablewidth{240pt}
\tablecaption{$10 < K < 13.5$ Galaxy Number Counts {\label{tab:ndensity}}} 
\tablehead{
\colhead{Region} & \colhead{$N$} & \colhead{Area (deg$^2$)} &
 \colhead{$n_{\rm gal} ({\rm deg}^{-2})$}  }
\startdata
XSC $|b| \ge 30$   & 363803   &   20630      &   17.63 \\
2dFGRS   &   32568    &    1887      &    17.26 \\
Sloan EDR     &    7078     &   369.6     &    19.15 \\
\enddata
\end{deluxetable}

\begin{figure*}[tb]
\center{\includegraphics[scale=1.0, angle=0]{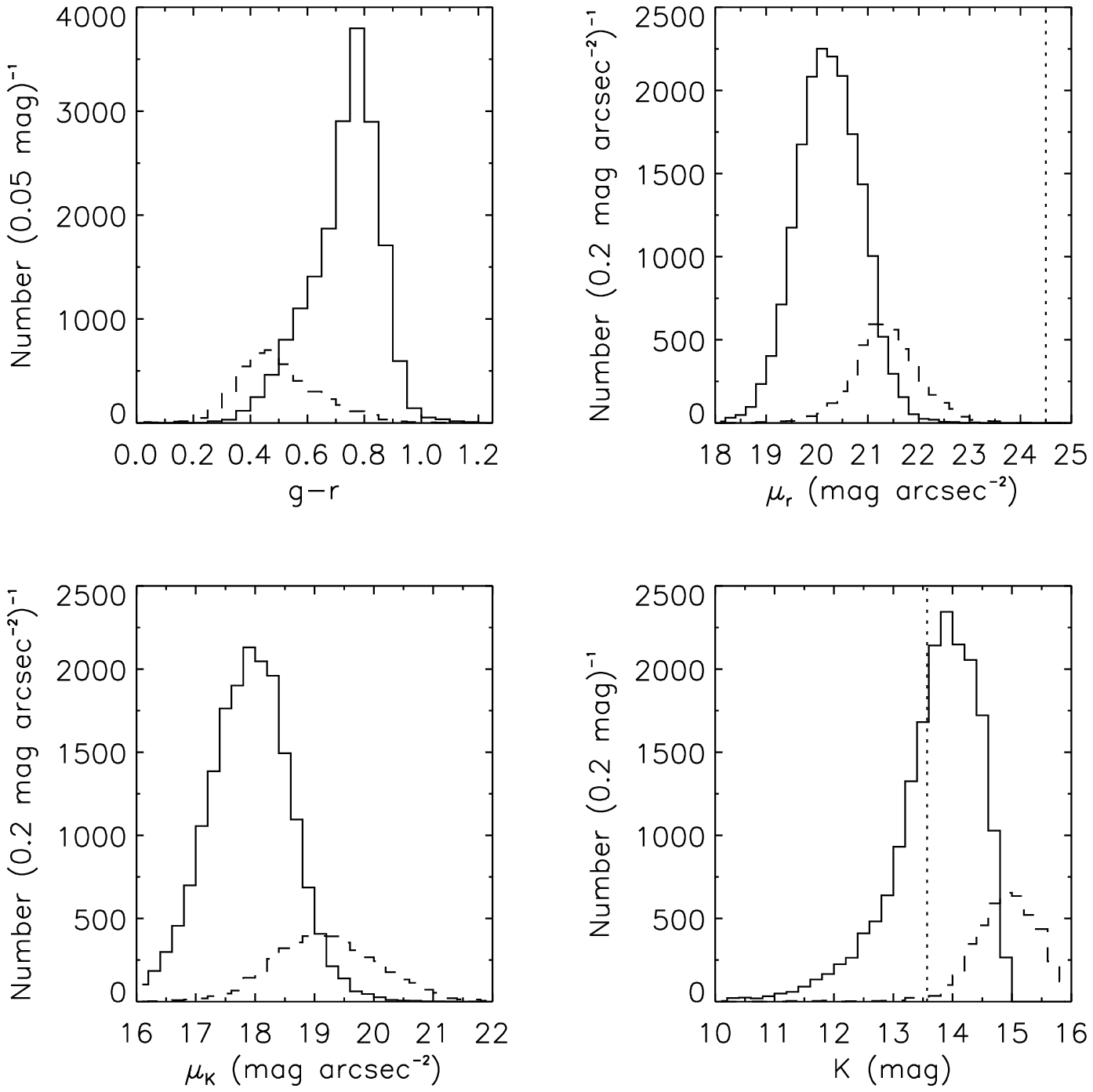}}
\caption{\label{fig:bluemissing} The $g-r$
colors, $r$ and $K$-band surface brightnesses, and 
$K$-band apparent magnitudes of 22679 SDSS $13 \le r \le 17.5$ galaxies
with ({\it solid}) and without ({\it dashed}) 2MASS matches.  
$K$-band surface brightnesses and apparent magnitudes
for the 3965 galaxies without 2MASS data are estimated using the
$r$-band derived quantities and the estimated $r-K$ 
color of the best-fit SED model, as described in \S
\ref{sec:method}. The dotted line in the upper right 
panel shows the surface brightness limit of SDSS, and 
the dotted line in the lower right panel shows the 
$K<13.57$ magnitude limit that we adopt for our LF analysis.}
\end{figure*}

We focus our study on the 
2MASS matches to the $13 \le r \le 17.5$ SDSS catalog.  As stated
earlier, out of the 22679 $13 \le r \le 17.5$ galaxies in the EDR spectroscopic 
sample, we match 12085 galaxies in the 2MASS XSC and
6629 galaxies in the 2MASS PSC.  In Fig.\
\ref{fig:bluemissing}, we explore the properties of the 2MASS matched
and unmatched galaxies in more detail.  In the upper panels,
we show the distribution of galaxy 
$g-r$ color ({\it left}) and $r$-band surface brightness 
$\mu_r$ ({\it right}).
The solid histograms show galaxies with 2MASS counterparts,
the dashed histograms show those without.  We give the $K$-band
surface brightness $\mu_K$ ({\it left}) and apparent magnitude $K$
({\it right}) in the lower panels.
We estimate $\mu_K$ and $K$ for galaxies that
have no 2MASS data using the SDSS $\mu_r$ and $r$-band apparent
magnitude in conjunction with the $r-K$ color of the best-fit
SED model (as described in \S \ref{sec:method}).  
We test this procedure by using the optical data only
to predict the $K$-band magnitudes of the 12085 galaxies with $K$-band XSC
data.  We find
that this procedure is accurate to 0.4 mag RMS.
We see that the galaxies that are unmatched in 2MASS are
preferentially blue and LSB in the 
optical and NIR. 
There are 84 LSB
galaxies ($\langle \mu_K \rangle \sim 19.1$ mag\,arcsec$^{-2}$)
with estimated $K<13.57$, thus, there may be
a small population of LSB galaxies
missed by 2MASS.  Faint, LSB galaxies are visible only in the very 
nearest parts of an apparent magnitude-limited survey (such as 2MASS),
and therefore carry a large weight $1/V_{\rm max}$.
Therefore, this small bias ($\sim 1$\%) may 
translate into a larger bias when considering the LF or luminosity density.
This bias would affect all published 2MASS LFs 
\citep[e.g.,][]{cole01,kochanek01},
as well as our own.  We show later that this bias
affects the faint end of the LF, as one would expect
given the surface brightness dependence of the LF 
\citep[see, e.g.,][]{dejong00,cross02}.
We also estimate the degree of incompleteness
using the optical data in conjunction with our knowledge of 
stellar populations to push the $K$-band LF and stellar MF down to 
lower galaxy masses.

We select samples for estimating LFs in different passbands using
passband-dependent magnitude limits, following \citet{blanton03}.
Specifically, when constructing $ugizK$ LFs, we select the magnitude limit
in $ugizK$ so that the $V_{\rm max}$ for each galaxy is 
constrained by the $ugizK$ limit for 98\% of the sample, and 
is defined by the $r\le17.5$ limit for the other 2\% of the 
galaxies.  Functionally, these limits are $u =$18.50, 
$g=$17.74, $i=$16.94, $z=$16.59, and $K=$13.57.

\section{Methodology: $k$-Corrections, Evolution Corrections, and Stellar M/L ratios}
\label{sec:method}

\subsection{The Method}

To estimate LFs and stellar MFs using
the redshift and $ugrizK$ data for the SDSS EDR galaxies, we must estimate
$k$-corrections and stellar M/L ratios.  Furthermore, 
\citet{blanton03} and \citet{norberg02} stress the need 
to include the effects of galaxy evolution.
We estimate $k$-corrections, evolution
corrections, and galaxy stellar M/L ratios by comparing the $ugrizK$
galaxy fluxes with state-of-the art stellar population
synthesis (SPS) models.  

For each galaxy, we construct a grid of stellar 
populations with a range of metallicities and
star formation histories (SFHs) at {\it both} the 
real galaxy redshift and at redshift zero.
We use the \pegase model \citep[see][for a description of an
earlier version of the model]{fioc97}, choosing ten 
galaxy metallicities from 0.5\% to 
250\% solar.  The SFHs vary exponentially
with time $t$: $\psi = [\tau^{-1}(1-e^{-T_0/\tau})^{-1}] e^{-t/\tau},$
where $\psi$ is the star formation rate (SFR), $\tau$ is the
exponential $e$-folding time of the SFR, and 
$T_0$ is the age of the galaxy (the time since SF commenced).
The term in the square brackets is simply a normalization to keep the 
total mass of stars formed by the present day 
at one solar mass.  We choose a grid of 29 $\tau$ values between 0 (single
burst) and $\infty$ (continuous), continuing through to 
$- \infty$ and then to $-1$\,Gyr (strongly increasing to the present day).
Our grid covers color space relatively uniformly.

This grid is produced at both redshift zero and 
at the real galaxy redshift.  The galaxy age
is 12\,Gyr for the redshift zero model, and is younger
for the non-zero redshift model assuming $h = 0.7$ (in essence,
we choose a formation redshift of $\sim 4$).  
For example, this gives an age of 10.7\,Gyr 
for a galaxy at $z=0.1$.  We least-squares fit the model
galaxies at the real galaxy redshift to the observed galaxy
colors to choose the best model galaxy template.  We then 
estimate the evolution 
correction, $k$-correction, and present-day stellar M/L ratio by 
comparing the non-zero redshift model with the evolved
redshift zero model.  Thus, in essence, we correct for 
evolution by assuming that the SFH
indicated by the colors of the galaxy at the observed redshift 
continues smoothly to the present day.

To estimate stellar masses, we adopt the $z=0$
model galaxy M/L ratios in each passband, assuming 
solar absolute magnitudes of (6.41, 5.15, 4.67, 4.56, 4.53, 3.32) 
in $ugrizK$ respectively, estimated using the \pegase SPS model.
Those wishing to convert our luminosity densities into physical
units or SDSS or 2MASS-calibrated absolute magnitudes per cubic Mpc can
easily use the above solar absolute magnitudes for conversion without
loss of accuracy.   
We estimate uncertainties
in $k$-corrections, evolutionary corrections, and stellar
M/L ratio values via three methods: ({\it i}) omitting one passband at a time
from the SED fit (the jackknife method; 6 fits); 
({\it ii}) uniformly weighting all data points in 
the fit for each galaxy (1 fit); and ({\it iii}) adding random
magnitude offsets with sizes corresponding to the magnitude error
to all the galaxy photometry and re-doing the fits (5 times).  
We then compute the errors
from the RMS difference between these 12 different
fits to the SED and our original SED fit.  Typical 
$k$-correction and evolution correction random errors derived in this way
for the $g$-band selected
sample are (0.06, 0.03, 0.03, 0.03, 0.03, 0.02) mag in $ugrizK$ respectively.
Typical random errors in stellar M/L ratio are 
(0.08, 0.06, 0.05, 0.04, 0.04, 0.05) dex, again in $ugrizK$.

We adopt a `diet' \citet{salp} 
stellar IMF \citep[following][]{ml}
that has the same colors and luminosity as a normal
Salpeter IMF, but with only 70\% of the mass due
to a lower number of faint low-mass stars.
This yields stellar M/L ratios 30\% lower at a given
color than a Salpeter IMF.
\citet{ml} show that this IMF is `maximum disk', inasmuch
as IMFs richer in low-mass stars over-predict the rotation velocity
of Ursa Major Cluster galaxies with $K$-band photometry and
well-resolved \hi rotation curves.  This prescription 
thus gives the maximum possible stellar M/L ratio.  Naturally,
a different choice of IMF allows lower M/L ratios.  For example, 
the popular \citet{kennicutt83} or \citet{kroupa93} IMFs 
have roughly 30\% lower M/L ratios than
this IMF, and are thus `submaximal'.  We discuss this 
issue in more detail in \S \ref{sec:imf}.

\subsection{Comparison with Other Constraints}

\begin{figure}[tb]
\vspace{-0.5cm}
\hspace{-0.5cm}
\epsfbox{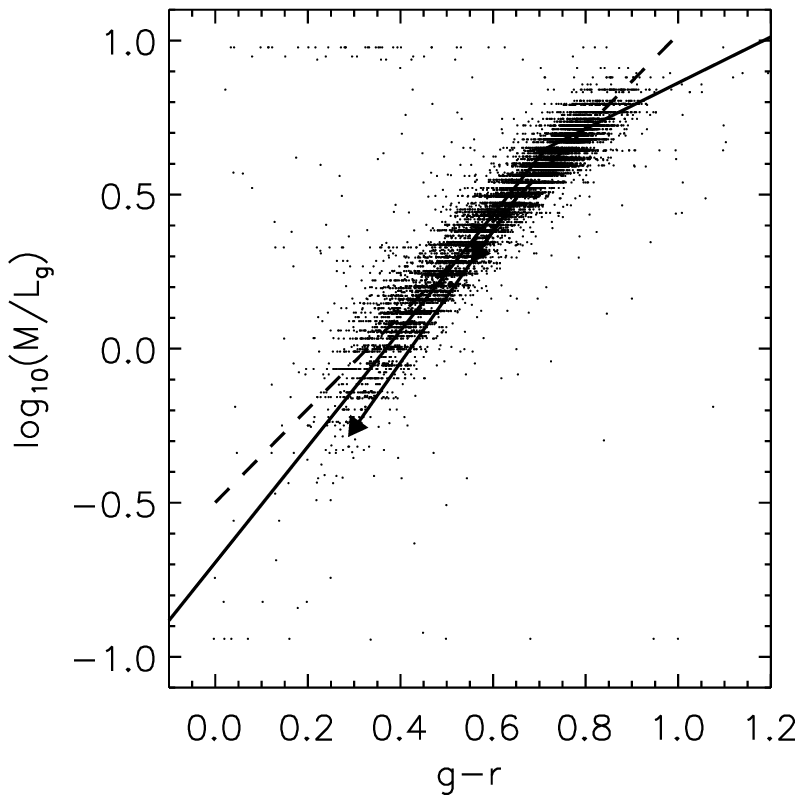}
\vspace{-0.2cm}
\caption{\label{fig:ml} Our $g$-band stellar
M/L ratio estimate (from the maximum-likelihood fit to the 
galaxy SED) against $g-r$ color for the 
$g$-band selected sample of 11848 galaxies.  The dashed
line is the bi-weight least-squares fit to the data. The solid
line is a rough fit to the relationship in Fig.\ 19 of 
\protect\citet{kauffmann03a},  accounting for the 
0.15 dex offset between a Kennicutt IMF and our diet
Salpeter IMF.
In addition, we transform the \protect\citet{kauffmann03a}
$z=0.1$ color to a $z=0$ color using $(g-r)^{z=0.1} \sim 0.91 (g - r)$
(see text for full details).  The
arrows show the average ({\it short arrow}) and maximum ({\it long arrow})
effect of dust on this relationship, as discussed in
\S \ref{sec:sysml}.
 }
\end{figure}

Our $k$-corrections and evolution corrections are quite robust.
$K$-band $k$-corrections are 
insensitive to galaxy spectral type. In particular, we find
$k(z) \sim -2.1 \pm 0.3 z$, which is in good agreement with
$k(z) \sim -2.25 z$ from \citet{glazebrook95}.
We test the optical $k$-corrections by comparing with a
simple power-law interpolation, including the effects of bandpass widening.
\citet{blanton03k} find
that this approximation is good to around 0.1 mag in all passbands, but 
better in $riz$ as the spectral shapes are simpler
there.  We find also that our optical and NIR $k$-corrections 
are consistent with 
the simple power-law recipe to within 0.1 mag in all
passbands.  These offsets decrease
to 0.05 mag in $riz$.  This agreement is more than adequate,
bearing in mind our 0.05--0.1 mag $k$-correction
errors.  We quantify
our evolution corrections
by comparing the mean $k+$evolution correction with the 
mean $k$-correction for our $g$-band selected galaxy sample.
The mean evolution corrections are $\sim (2.3, 1.6, 1.3, 1.1, 1.0, 0.8)z$ in
$ugrizK$, in the sense that galaxies are fainter at the present
day, owing mostly to passive evolution.
This can be compared to the $Q$ values
derived by \citet{blanton03}, who estimate the evolution by 
fitting for it explicitly in their LF estimation.  They find an
evolution of $\sim (4.2\pm0.9, 2.0\pm0.5, 1.6\pm0.3, 1.6\pm0.4, 0.8\pm0.3)z$
in $ugriz$.  Therefore, we find satisfactory agreement between
our color-based evolutionary corrections and direct estimates
from the LF evolution by \citet{blanton03}, except perhaps in the $u$-band,
where the photometric and $k$-correction uncertainties are largest, 
and our assumption of smoothly-varying SFHs could
easily prove inadequate.  Independently, \citet{bernardi_ii} find evolution
of $\sim (1.2, 0.9, 0.8, 0.6)z$ in $griz$ for early-type 
galaxies using a similar (but totally independent)
technique to \citet{blanton03}, again
within $\la 0.05$ mag of our corrections 
over the redshift interval of interest.

\citet{ml} demonstrate that for galaxies with relatively smooth SFHs, 
stellar M/L ratio and optical color should correlate quite tightly.
We present a test of our stellar M/L ratio estimates
in Fig.\ \ref{fig:ml}.  Using an
independent method that accounts for bursts of SF
based on the strengths of the 4000{\AA} break
and the H$\delta$ line,
\citet{kauffmann03a} 
construct stellar M/L ratios for over 120,000 SDSS galaxies.  In Fig.\
19 of that paper, they compare their M/L ratios in $g$-band
with the $g-r$ color, estimated at $z=0.1$, and find a strong correlation.
To compare to the \citet{kauffmann03a} correlation we
estimate a color correction $(g-r)^{z=0.1} \sim 0.91 (g - r)$,
assuming a power-law $k$-correction.  Moreover, we account
for the IMF difference; our `diet' Salpeter
IMF is 0.15 dex heavier at a given color, because of its larger number of 
low-mass stars, than the Kennicutt IMF that 
\citet{kauffmann03a} adopt.  To within 20\% random scatter
our multi-color method gives results consistent
with their spectral method (comparing the points with 
the solid line in Fig.\ \ref{fig:ml}). This is particularly impressive given
the very different methodologies and the different stellar population 
models used\footnote{\citet{lee03_2} show
a comparison of their color-derived two-population 
$B$-band stellar M/L ratio estimates \citep{lee03} with optical color
in the Appendix to that paper.  They find excellent 
agreement with color-derived stellar M/L ratios from 
\citet{ml}, showing again that different
methodologies yield consistent estimates of stellar
M/L ratio.}.  

With the low scatter in the 
$g-r$ versus M/L$_g$ ratio correlation it is possible
to predict stellar mass to within 20\% using $g$- and $r$-band data alone,
compared to the maximum-likelihood SED fits of 
up to 6 optical/NIR passbands.  \citet{kauffmann03a} 
find a scatter closer to 50\%; this is likely due to the 
different methods adopted to derive stellar M/L ratios by our group and 
\citet{kauffmann03a}.  \citet{kauffmann03a} use
3{\arcsec} aperture spectra, plus an
$r-i$ color that is emission-line sensitive for dust estimation.  Thus, they
are sensitive to the aperture mismatch between the spectra and colors,
and model mismatches between color and spectral features.  In contrast,
we minimize the residuals explicitly between our galaxy model
colors and the observed colors; therefore, we explicitly 
minimize the spread in the color--M/L ratio correlation 
with our method.  Either way, it is clear
that we can use SDSS color
data alone, plus a redshift, to estimate the stellar mass of galaxies to 
between 20\% and 50\%, relative to the answer that one obtains using
$K$-band data or spectra.  In particular, 
this allows us to use the SDSS data to `fill in'
areas of parameter space not covered as completely by 2MASS, such as
blue LSB galaxies.  

We can independently check these M/L ratios using recent results
from \citet{bernardi_iii}.  They construct estimates of 
total M/L ratio (including the potentially non-negligible contribution of
dark matter) using kinematic constraints, by multiplying 
the half-light radius by the velocity dispersion squared, and then dividing
by half the luminosity.  Assuming the {\it Hubble Space Telescope}
Key Project distance scale \citep{freedman01}, they then compare
these M/L ratios with $g-r$ color, finding a strong correlation 
(the lower right-hand panel of their Fig.\ 5).  Correcting
for their application of a 0.08 mag bluewards offset in $g-r$ color, 
we estimate their $\log_{10} (M/L_r) \sim -0.15 + 0.93 (g-r)$.  Over the 
$g-r$ range of interest ($0.3 \la g-r \la 1$), this 
is within 25\% at the blue end and 5\% of the red end
of our maximum-disk tuned stellar population 
model expectation (see the Appendix for details):
$\log_{10} (M/L_r) = -0.306 + 1.097 (g-r)$.  Furthermore, 
their total scatter (including contributions from observational
error) is $\sim 0.15$ dex, or 40\% in terms of M/L ratio, in agreement
with our earlier estimate of 20\% to 50\%.  The agreement between 
these two totally independent methodologies, each with their
own sources of systematic and random error, is astonishing; 
both predict roughly a factor of five change in stellar M/L ratio
from the blue to the red end of the galaxy population, and both
have the same absolute stellar M/L ratio scale.  This agreement 
is another, powerful argument in favor of a color-based
stellar M/L ratio of the type discussed in the Appendix, or 
by \citet{ml}. 

\subsection{Systematic uncertainties} \label{sec:sysml}

The above prescription for estimating $k$-corrections, evolution 
corrections, and stellar M/L ratios assumes that the colors of a stellar
population are driven by star formation history (SFH) and metallicity alone.
What are the systematic uncertainties introduced by neglecting 
the effects of dust and more complex SFHs?  Our $k$-corrections are 
robust, inasmuch as we simply use a physically-motivated 
model to interpolate between the observations (a simple power-law
interpolation suffices also to roughly 0.1 mag).  Furthermore, our
evolutionary corrections agree with independent estimates, 
and since they are a relatively
small correction ($\la 0.2$ mag typically), small errors in 
the evolutionary correction will not substantially affect our 
results.  

However, there may be a significant uncertainty in stellar M/L ratio estimates
that is not accounted for by our prescription.  Overall
galaxy age (i.e.\ the time since SF started) can
change the stellar M/L ratio at a given color in a systematic sense
by a small but non-negligible amount, e.g.\ $\pm 0.05$ dex for
an age difference of $\pm 3$\,Gyr.  Furthermore, it is 
not {\it a priori} clear what effects dust
may have on the stellar M/L ratios.  \citet{ml} show that, 
to first order, the effects of dust cancel out to within 
0.1--0.2 dex when estimating
color-derived stellar M/L ratios.  This cancellation occurs because the stellar
populations and dust each predict roughly the same amount of reddening
per unit fading in most passbands.  However, the random uncertainties
of this technique are only 20\% in terms of M/L ratio (see above earlier
in \S \ref{sec:method}), 
so the second order difference between the effects of dust and stellar
populations could be significant.

We explore the possible effects of dust on our results in 
a simple way, following \citet{tully98}.  
\citet{tully98} estimate the
luminosity-dependent dust content of disk galaxies by minimizing
the scatter in the color-magnitude relation (CMR) in $BRIK$ passbands.
They find that luminous galaxies suffer from a 1.7 (0.3) mag 
dimming in their $B$-band ($K$-band) flux when going from face-on to nearly
edge-on, while faint galaxies show very little evidence for dust.  
We adopt a rough dimming of (1.6, 1.3, 0.3) mag from face-on
to edge-on ($\ge 80\arcdeg$) in $grK$ passbands for massive disk galaxies 
(masses $> 3 \times 10^{10} h^{-2} M_{\sun}$ and $c_r < 2.6$).  We allow
this dimming to decrease linearly with logarithmic mass to zero for stellar
masses below $3 \times 10^8 h^{-2} M_{\sun}$.  We assume a simple 
slab model, with an optical depth at arbitrary inclination of
$\tau(i) = \tau(80\arcdeg) \cos 80\arcdeg/\cos i$, where
$\tau(80\arcdeg)$ is the optical depth at edge-on derived from
the above quoted difference between edge-on and face-on.
Concentrated, i.e.\ early-type, galaxies are assigned 
a factor of 3 less dust.  We assume a random distribution of 
orientations. 

In Fig.\ \ref{fig:ml}, we show schematically 
the effect of the average ({\it short arrow}) and maximum
possible ({\it long arrow}) dust contents, according to our admittedly
{\it ad hoc} description.  The arrows show the bluing of color
and reduction of M/L ratio when dust is taken into account.  It is clear
that the effects of dust and stellar
population are mostly degenerate in agreement with \citet{ml}.  
Nevertheless, 
there is a slight
systematic difference between the two effects.  For
this dust prescription, we overestimate the average stellar M/L ratio
in $g$-band by 0.06 dex when we fit dust-reddened colors with pure
stellar populations.   

Another source of systematic uncertainty is from bursts of SF.
\citet{ml} find that large bursts of 
SF can cause an over-estimate of the true M/L ratio
of the stellar population, if the stellar population is interpreted
in terms of smoothly varying SFHs.  We attempt to constrain
the magnitude of this error for our purposes using a simple model.
We choose two solar metallicity stellar populations, one with 
a decreasing SFR to the present day ($\tau = 4$\,Gyr), and one with constant
SF ($\tau = \infty$\,Gyr).  We then apply random 
variations in SFR over timescales of 10$^8$\,yr, distributed in a 
log-normal fashion with a dispersion $\sigma$
of a dex, i.e. the SFR can easily change by more than an order of
magnitude from its baseline rate.  We then examine the offset
from the $g-r$ color and M/L$_g$ ratio correlation of these bursty
models, compared to smooth SFH models.  For both SF models we find
that SF bursts generate a
$\sim 25$\% scatter about the color--M/L ratio relation, 
and a $\sim 10$\% offset to slightly lower
M/L ratio at a given color.  A full order of magnitude variation
in SFH over 10$^8$\,yr timescales is likely to be an upper
limit for all but the strongest present-day star-bursting galaxies;
therefore, we demonstrate that the bias we impose by assuming such simplistic
SFHs is $\la 10$\%.

To summarize, the random uncertainties of color-based stellar M/L ratio
estimation are $\sim 20$\%.  Systematic uncertainties from 
galaxy age, dust, and bursts of SF are $\sim 0.1$ dex, 
or $\sim 25$\%.  These systematic uncertainties will not cancel
out with larger galaxy samples, and will 
dominate, along with stellar IMF, the
error budget of the stellar mass density of the Universe.

\section{Luminosity Functions} \label{sec:lf}

\begin{table*}
\caption{Systematic Error Budget {\label{tab:sys}}}
\begin{center}
\begin{tabular}{lcll}
\hline
\hline
Quantity & Error & Source & Ref. \\
(1) & (2) & (3) & (4) \\
\hline
\multicolumn{4}{c}{Luminosity Function} \\
\hline
$\phi^*$ & 10\% & Uncertainty in exact sky coverage (3\%), completeness (7\%),
	Poisson error in  normalization & 
	\S \ref{samp:compl}, \S \ref{sec:lfest} \\ 
& & (1\%), and differences between behavior of the $10 < K < 13.5$ sample 
	and our EDR sample & \\
$M^*$ & 5\% & Uncertainty in absolute calibration 
	of $ugrizK$ system & (1) \\
& 10\% & {\it $K$ only:} Extrapolation to total & \S \ref{magacc} \\
$\alpha$ & 0.1? & {\it Optical:} from departures 
	from a Schechter function & \S \ref{gband:samp} \\
& $^{+0.1}_{-0.6}$ & {\it NIR:} from strong departures from a 
	Schechter function, and LSB galaxy incompleteness 
	& \S \ref{samp:compl}, \S \ref{sec:k} \\
$j$ & 15\% & {\it Optical:} from $\phi^*$ and $M^*$ uncertainty 
	& above \\
& $ ^{+35\%}_{-15\%}$ & {\it NIR:} from $\phi^*$, $M^*$ and $\alpha$ 
	uncertainty & above \& \S \ref{sec:k} \\

\hline
\multicolumn{4}{c}{Stellar Mass Function} \\
\hline
$M^*$ \& $\rho$ & 30\% & Dust, bursts of SF, galaxy age, and 
	absolute calibration uncertainty & above \& \S \ref{sec:sysml} \\
& $ ^{+0\%}_{-60\%}$ & Stellar IMF & \S \ref{sec:imf} \\

\hline
\end{tabular}
\\ 
\vspace{-1.0cm}
\tablerefs{ (1) \citet{fuku96} }
\vspace{-0.4cm}
\tablecomments{Column (1) describes the quantity, (2) the contribution
to the systematic error budget, (3) describes the error in more detail, 
and (4) gives any relevant references (section number or literature
citation).
}
\end{center}
\end{table*}

\begin{table*}
\caption{Galaxy Luminosity Function Fits {\label{tab:fits}}}
\begin{center}
\begin{tabular}{lcccccccccc}
\hline
\hline
Band & $m_{\rm lim}$ & $N_{\rm gal}$ & $\langle V/V_{\rm max} \rangle$ &
$\langle z \rangle$ & $\phi^*$ & $M^* - 5 \log_{10} h$ &
$\alpha$ & $j$ & $j_{\rm literature}$ & Ref. \\
(1) & (2) & (3) & (4) & (5) & (6) & (7) & (8) & (9) & (10) \\
\hline
$u$ & 18.50 & 5347 & 0.532$\pm$0.004 & 0.055 & 0.0238(8) 
& $-$18.13(3) & $-$0.95(3) & 
1.51$^{+0.03}_{-0.04}\times 10^8$ & 1.45$\times 10^8$ & a \\
$g$ & 17.74 & 11848 & 0.509$\pm$0.003 & 0.070 & 
0.0172(5) & $-$19.73(3) & $-$1.03(3) & 
1.57$^{+0.02}_{-0.06}\times 10^8$ & 1.47$\times 10^8$ & a \\
$r$ & 17.50 & 22679 & 0.509$\pm$0.002 & 0.096 & 0.0137(7) 
& $-$20.57(3) & $-$1.07(3) &
1.80$^{+0.03}_{-0.08}\times 10^8$ & 1.69$\times 10^8$ & a \\
$i$ & 16.94 & 17984 & 0.508$\pm$0.002 & 0.093 & 0.0118(4) 
& $-$21.00(3) & $-$1.11(3) & 
2.14$^{+0.02}_{-0.13}\times 10^8$ & 2.19$\times 10^8$ & a \\
$z$ & 16.59 & 15958 & 0.520$\pm$0.002 & 0.092 & 0.0119(4) 
& $-$21.34(2) & $-$1.06(2) &
2.75$^{+0.03}_{-0.14}\times 10^8$ & 3.22$\times 10^8$ & a \\
$K$ & 13.57 & 6282 & 0.520$\pm$ 0.004 & 0.078 & 0.0143(7) & $-$23.29(5) & $-$0.77(4) 
& $5.8^{+1.8}_{-0.1}\times10^8$ & 5.9,7$\times10^8$ & b,c \\
\hline
\end{tabular}
\\ 
\vspace{-1.0cm}
\tablecomments{The passband (1), corresponding limiting magnitude (2), and
number of galaxies (3).  The mean $V/V_{\rm max}$ is in (4), and 
the mean redshift in (5).  Each passband
LF is fit with a Schechter function and described by three parameters --
the normalization $\phi^*$ in $h^3$\,Mpc$^{-3}\,{\rm mag}^{-1}$ (6),
the characteristic luminosity $L^*$ (7), and the faint end slope $\alpha$ (8).
Our luminosity density estimate $j$
(9), compared to an estimate from the literature (10), both in units
of $h\,L_{\sun}\,{\rm Mpc}^{-3}$.  The literature
references are as follows: (a) \citet{blanton03}; (b) \citet{cole01}; 
(c) \citet{kochanek01}.  The formal error estimates for quantities are given
in parentheses.
Table \ref{tab:sys} gives a complete summary of the systematic
error sources, in addition to the formal errors 
calculated above.  The optical luminosity densities give the 
formal error as the positive error bar, and the influence of 
the correction of early-type galaxy magnitudes by $-$0.1 mag
in SDSS as the negative error bar.
The $K$-band error estimate includes a substantial uncertainty
from 2MASS's bias against LSB galaxies.  
See Fig.\ \ref{fig:lumdens} for a graphical 
representation of the luminosity density literature comparison.  }
\end{center}
\end{table*}

\subsection{LF estimation} \label{sec:lfest}

We estimate LFs using the simple and intuitive $V/V_{\rm max}$ 
formalism of, e.g., \citet{felten}.  This method has the disadvantage that 
it is somewhat sensitive to galaxy density variations.
For example, if the near part of the survey is rather overdense, 
where a magnitude-limited
survey is most sensitive to low-luminosity galaxies,
then the $V/V_{\rm max}$ estimator will yield a somewhat larger
number of low-luminosity galaxies than it should.
Both the Step-Wise Maximum Likelihood (SWML) method of \citet{eep} 
and the parametric method of \citet{sty} are insensitive to
density fluctuations of this type (although both methods are sensitive to 
density fluctuations when calculating the overall LF normalization).
Nevertheless, both the SWML and parametric method make the assumption
that the shape of the LF is independent of environment, yet there
is impressive evidence against this assumption, at least 
in the optical \citep{deprop03,hutsi03}.
In contrast, the $V/V_{\rm max}$ method does not make this assumption.
Furthermore, it does not make any {\it a priori} assumptions 
regarding the form of the LF, unlike \citet{sty} parametric method.
Therefore, we use the $V/V_{\rm max}$ method and note, that among others,
\citet{cole01} find with a similar dataset that LFs 
derived using $V/V_{\rm max}$ and SWML are identical
within the errors.

For an unbiased estimate of $V_{\rm max}$, we estimate the 
maximum distance that a galaxy of a given absolute magnitude
would be visible, accounting for Galactic foreground extinction and
$k$- and evolution-corrections,
not including the early-type and $K$-band-to-total
corrections.  Due to slight inaccuracies in the $k$- and evolution corrections,
we find a few galaxies with distances
that are larger by a few observational sigma
than expected, which gives a few
galaxies with $V/V_{\rm max} > 1$.  Not including these
galaxies does not affect any of the results in this
paper.  Our formal error estimates include Poisson, Monte-Carlo
magnitude, evolution, $k$-correction, and $V/V_{\rm max}$
bootstrap uncertainties, plus 
random stellar M/L ratio errors for the MFs.  There are also systematic
sources of error: e.g., the $\sim 25$\% 
systematic uncertainty from dust and bursts of SF, and the 
$\sim 5$\% systematic uncertainty in absolute magnitude
calibration \citep[see, e.g.,][]{fuku96}.  
We have to first order avoided uncertainties from
galaxy clustering because we have renormalized the luminosity
functions to account for the EDR's 8\% overdensity (\S \ref{samp:compl}).  
However, we neither sample all of the EDR region, nor can we securely
extrapolate to different galaxy populations (e.g., the $13\le r \le17.5$ 
population as opposed to the $10 \le K \le 13.5$ population), so
we attach a 10\% percent systematic uncertainty to the normalization
$\phi^*$ and luminosity density $j$
to account for clustering.  We summarize the systematic error budget in 
Table \ref{tab:sys}.

We calculate LFs using pseudo-$ugrizK$-limited
samples (where 98\% of the galaxies are limited in the passband
of interest, and only 2\% of the galaxies are limited by 
their $r$-band flux).  
We present our results in Table \ref{tab:fits} and include some
relevant comparisons from the literature.
We discuss two sets of LFs in more detail in this paper: a 
joint $r$ and $K$-band selected sample, and a 
joint $g$ and $r$-band selected sample such that only
2\% of galaxies are $r$-band limited, and 98\% are
limited by the magnitude limit in the other passband.
We do not calculate the LF or stellar MF 
for magnitude bins with less than 5 galaxies.

\subsection{The $K$-band limited sample} \label{sec:k}

\begin{figure}[tb]
\epsfxsize=8.5cm
\epsfbox{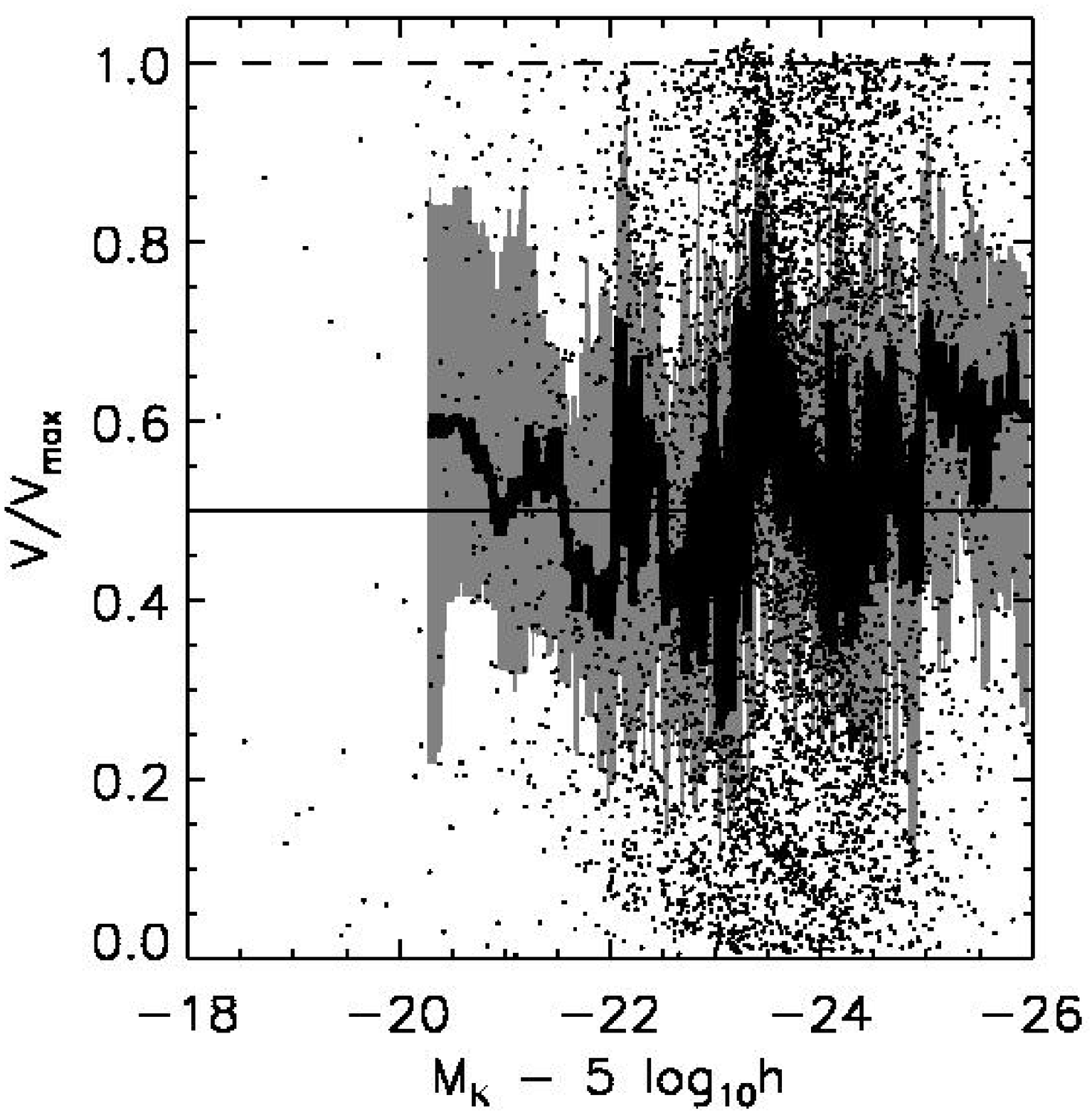}
\caption{\label{fig:comp} $V/V_{\rm max}$ versus $K$-band 
absolute magnitude.  The median ({\it thick solid line}), and 
upper and lower quartiles ({\it shaded area}), are shown as a function
of $K$-band absolute magnitude.  The 
average value for the whole sample is $0.520\pm 0.004$, which is reasonably
consistent with the expected value of 0.5 ({\it thin solid line}).  }
\end{figure}

\begin{figure}[tb]
\vspace{-0.5cm}
\hspace{-0.5cm}
\epsfbox{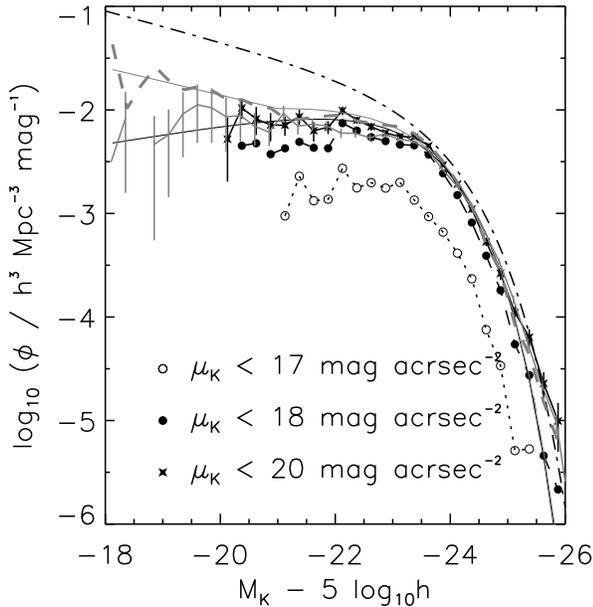}
\vspace{-0.2cm}
\caption{\label{fig:kbandtotal} $K$-band $V/V_{\rm max}$ LFs using
different surface brightness cuts.
The black solid line with data points represents
the total sample ($\mu_K < 20$\,mag\,arcsec$^{-2}$) LF.  The dotted and dashed
lines represent the LF for $\mu_K < 17$\,mag\,arcsec$^{-2}$ and 
$\mu_K < 18$\,mag\,arcsec$^{-2}$ subsamples, which
shows the LF steepening at the faint end as the 
surface brightness limit gets fainter.  The thick grey dashed
line denotes the predicted $K$-band LF, in the absence of selection 
bias (see the text for more details).  The thin grey line
shows the hybrid Schechter$+$power-law fit to the predicted $K$-band LF.
The grey solid line with error bars
denotes the 2MASS$+$2dFGRS LF of 
\protect\citep{cole01}, and the thin solid line is the Schechter fit
to our total LF, described in Table \ref{tab:fits}. 
For reference, the Schechter function fit to the 
total \citet{huang03} $K$-band LF is 
shown as a dash-dotted curve.  }
\end{figure}

\begin{figure}[tb]
\vspace{-0.5cm}
\hspace{-0.5cm}
\epsfbox{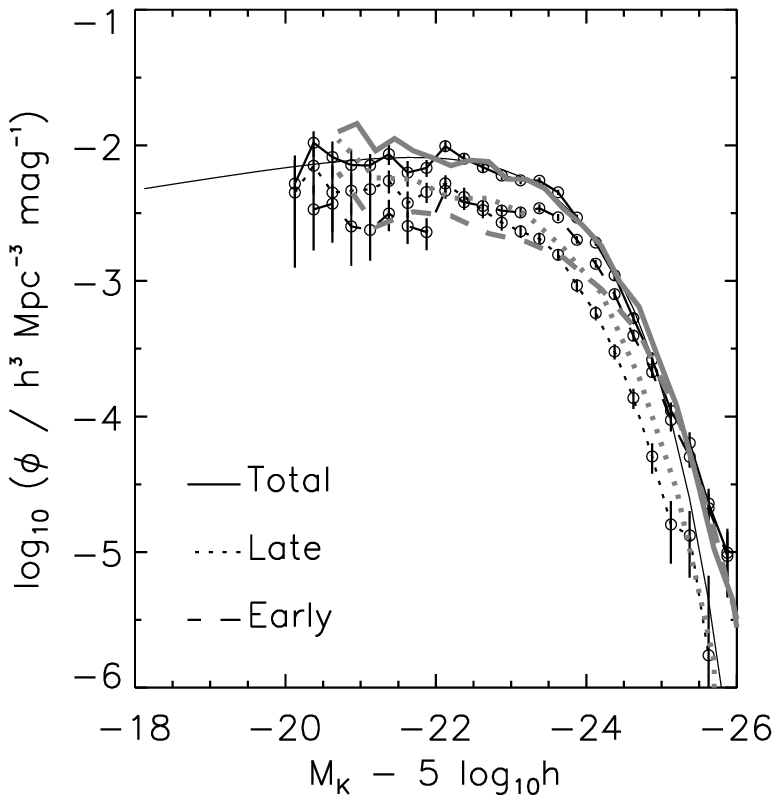}
\vspace{-0.2cm}
\caption{\label{fig:kbandsplit} $K$-band LF
split by morphological type.  
The solid line represents
the total LF.  The dotted and dashed
lines represent the LF for late and early-type
galaxies, separated using
$c_r=2.6$.  The solid grey line 
denotes the 2MASS LF of 
\protect\citep{kochanek01}, while the grey thick dashed and dotted lines
denote the early/late types from that paper, respectively. }
\end{figure}

In Fig.\ \ref{fig:comp}, we plot $V/V_{\rm max}$ versus $K$-band
absolute magnitude.  In an unbiased sample, an average value of 
$V/V_{\rm max} = 0.5$ is expected, as galaxies uniformly fill the volume.
For our sample, the average value is $0.520\pm 0.004$.  
Excluding the 66/6282 (1\%) of
galaxies with $V/V_{\rm max} > 1$ gives 
$\langle V/V_{\rm max} \rangle = 0.515$.  This indicates
a slight tendency for galaxies to be in the more 
distant half of the sample, perhaps reflecting 
uncertainty in the evolution correction
or small amounts of large-scale structure.  Nevertheless,
any bias in the sample is weak; for example, \citet{cole01}
find $\langle V/V_{\rm max} \rangle \simeq 0.52$ for their sample
of 2MASS/2dFGRS galaxies, yet obtain
excellent agreement between $V/V_{\rm max}$
and SWML estimates of the LF.

We plot the LFs derived using the $K$-band limited sample in
Figs. \ref{fig:kbandtotal} and \ref{fig:kbandsplit}.  
In Fig.\ \ref{fig:kbandtotal}, we show the 
$K$-band LF for samples using different $K$-band surface brightness
$\mu_K$ cuts.  All galaxies in our sample ({\it solid line}) have 
$\mu_K<20$\,mag\,arcsec$^{-2}$.  We fit the \citet{schechter} function
to the $V/V_{\rm max}$ data points:
\begin{equation}  
\phi(L){\rm d}L = \phi^* \left( \frac{L}{L^*} \right)^{\alpha} \exp{\left( -\frac{L}{L^*} \right)} \frac{{\rm d}L}{L^*} , 
\end{equation}
where $\phi^*$ is the LF normalization, $L^*$ is the characteristic
luminosity at the `knee' of the LF where the form changes from exponential
to power law, and $\alpha$ is the `faint end slope'.  In common
with other work \citep[e.g.,][]{dejong00,cross02},
we find that a fainter surface brightness limit increases
$\phi^*$ somewhat and substantially affects $\alpha$.
The $K$-band luminosity
density for our galaxy sample is 
5.77$\pm0.13\times10^8\,h\,L_{\sun}$\,Mpc$^{-3}$ (formal error only).
As discussed in Table \ref{tab:sys}, we estimate
a $\pm 15$\% systematic uncertainty from our extrapolation 
to total flux, absolute magnitude calibration, and sky coverage
uncertainty.
Thus, our raw estimate of $K$-band luminosity
density is 5.8$\pm0.9\times10^8\,h\,L_{\sun}$\,Mpc$^{-3}$, 
including the sources of random and systematic error.

Earlier, we expressed concern regarding incompleteness 
in 2MASS for LSB galaxies.  It is interesting to use the full SDSS$+$2MASS
dataset to estimate what the $K$-band LF {\it should} look like, in 
the absence of selection bias.  We use the $g$-band selected
galaxy sample to construct a $K$-band LF using 9307 real 
$K$-band magnitudes and 2541 synthesized $K$-band magnitudes
(estimated to $\sim 0.4$ mag accuracy using $ugriz$ as
a constraint).  This is denoted in Fig.\ \ref{fig:kbandtotal} 
as a thick grey dashed line.  The agreement is excellent at the bright end;
nevertheless, the faint end slope
of the predicted $K$-band LF is substantially steeper.
A Schechter function is a poor fit to this LF owing to the `kink' at
$M_K - 5\log_{10}h \sim -21$, thus we fit a power law to 
the faint end between $-21 \le M_K - 5\log_{10}h \le -18$,
which has a slope of $\alpha \sim -1.33$.  This bias against faint,
LSB galaxies affects all 2MASS-derived estimates
of not just the faint end slope,
but also the total $K$-band luminosity density.  Using this
rough hybrid Schechter$+$power law, 
which has $(\phi^*,M^*,\alpha) = (0.0149,-23.33,-0.88)$ 
brightwards of 
$M_K - 5\log_{10}h = -21$ and continues with power
law slope $-1.33$ faintwards of this limit (the solid grey
line in Fig.\ \ref{fig:kbandtotal}),
we estimate that the total $K$-band 
luminosity density may be as high as 
7.6$\times10^8\,h\,L_{\sun}$\,Mpc$^{-3}$.  Thus, we see that 2MASS's bias
against LSB galaxies may bias the faint end slope downwards, and
the luminosity density estimates
downwards by 25\%.  This conclusion is qualitatively and quantitatively
consistent with a more direct assessment of light missed by
2MASS's relatively shallow exposures by \citet{andreon02}
\footnote{There is another argument that suggests that 2MASS misses
LSB galaxies.  The luminosity-density weighted $r-K$ color (AB-Vega) 
of the galaxy population is 2.75$\pm$0.05 (when either $r$ or 
$K$-band luminosity weighted), and is dominated by luminous, 
$\sim L^*$ galaxies.  SDSS should not miss large numbers of 
LSB galaxies, therefore we can use the $r$-band luminosity 
density (which should be quite complete) plus the luminosity
density-weighted $r-K$ color (which reflects the 
behavior of $\sim L^*$ galaxies) to estimate the $K$-band luminosity
density.  This estimate is 
$j_K \sim 6.5\times10^8\,h\,L_{\sun}\,{\rm Mpc}^{-3}$,
somewhat higher than the uncorrected value of 
$5.8\times10^8\,h\,L_{\sun}\,{\rm Mpc}^{-3}$.  The virtue of this
$r-(r-K)$-based estimate is that the faint end slope is determined
by the well-constrained $r$-band LF, whereas in the direct approach
we are forced to estimate the faint end slope from the $K$-band data 
directly.  }.

How do our 
luminosity functions and luminosity density estimates 
of 5.8$^{+2.9}_{-0.9}\times10^8\,h\,L_{\sun}$\,Mpc$^{-3}$ compare
with the literature?  
In Fig.\ \ref{fig:kbandtotal}
({\it solid grey line}) we compare with 
the LF estimate of \citet{cole01} and find
excellent agreement in both the shape of the LF and
the overall normalization.  
\citet{cole01} find 5.9$\pm0.9 \times10^8\,h\,L_{\sun}$\,Mpc$^{-3}$
but do not account for the bias against LSB galaxies inherent in the
2MASS data.  
We compare our LF with \citet{kochanek01} in Fig.\ \ref{fig:kbandsplit}
(shown as a solid grey line), finding that their LF is well within
our error bars.\footnote{They also do not account for 2MASS's bias against
LSB galaxies but still find a relatively steep faint
end slope to the LF, making it hard to estimate
the effect of the LSB bias in their case.}  
They find a somewhat steeper faint end slope than
we do, leading them to a slightly high luminosity density of 
$\sim 7\pm 1 \times10^8\,h\,L_{\sun}$\,Mpc$^{-3}$.  Therefore,
accounting for all of the sources of error, it is clear that
we are consistent with both \citet{cole01} and \citet{kochanek01}. 
Our determination has the advantage, however, that we have considerably
reduced the large-scale structure uncertainty by renormalizing our
LF to the whole $|b|>30\arcdeg$ sky, and that we understand
in detail 2MASS's bias against LSB galaxies.  It will indeed 
be interesting to see if large, deep $K$-band surveys will
converge towards the steeper faint end slope predicted by our analysis.

In this context, comparison to the rather deeper survey 
of \citet{huang03} is particularly interesting.  They have a
sample of $\sim 1000$ galaxies over an area of sky 50 times smaller
than our area, and are unable to normalize their luminosity
function to the whole $|b|>30\arcdeg$ sky.  Thus, their luminosity
function is highly susceptible to the effects of large-scale structure.
We disagree with their luminosity function (see Fig.\ \ref{fig:kbandtotal})
and total luminosity density of $\sim 12 \times10^8\,h\,L_{\sun}$\,Mpc$^{-3}$.
We attribute much of this mismatch to large-scale structure.  
Two other effects may also contribute.  
First, the knee of their
LF is $\sim0.2$ mag brighter than ours, which is likely caused by
the uncorrected galaxy evolution in their sample.  Given a median redshift of
$z\sim 0.15$, a $-0.12 \pm 0.05$ mag offset is needed to correct for this
evolution.  Furthermore, ignoring even modest evolution can cause
faint end slope over-estimation with the maximum-likelihood
SWML and STY methods \citep{blanton03}.
Second, a Schechter fit poorly represents their dataset;
an improved fit would have a sharper knee and a shallower
faint-end slope, giving a substantially lower luminosity density
\citep[see Fig.\ 2 of][]{huang03}.  However, it is intriguing
that with deeper data they find a steep faint end slope, roughly
parallel with our predicted $K$-band LF.  We predict that 
further work will show that \citet{huang03} indeed found 
roughly the right faint end slope, but that they were adversely
affected by large-scale structure and a small offset from 
ignoring evolution corrections.

In Fig.\ \ref{fig:kbandsplit}, we show the LF split
crudely by morphological type using the 
SDSS $r$-band concentration parameter (\S \ref{sec:cnc}).
Recall that \citet{strat} and \citet{hogg02} show that
most concentrated $c_r \ge 2.6$ galaxies are early-type
(earlier than Sa), although $c_r$ could be 
affected by seeing \citep[see][and \S2.3]{blanton03c}.
We recover the classic result that the LF for early types
has a flat or decreasing faint end slope
and has a brighter $L^*$ than late types, which
have a somewhat steeper LF
\citep[e.g.,][]{bromley98,blanton01}.
Our type-split LFs agree qualitatively
with \citet{kochanek01}, who find a larger 
$L^*$ for early types, although their overall LF
is slightly offset from ours.
We explore the role of morphological selection in \S \ref{sec:col}.

\subsection{The $g$-band limited sample} \label{gband:samp}

\begin{figure}[tb]
\epsfxsize=8.5cm
\epsfbox{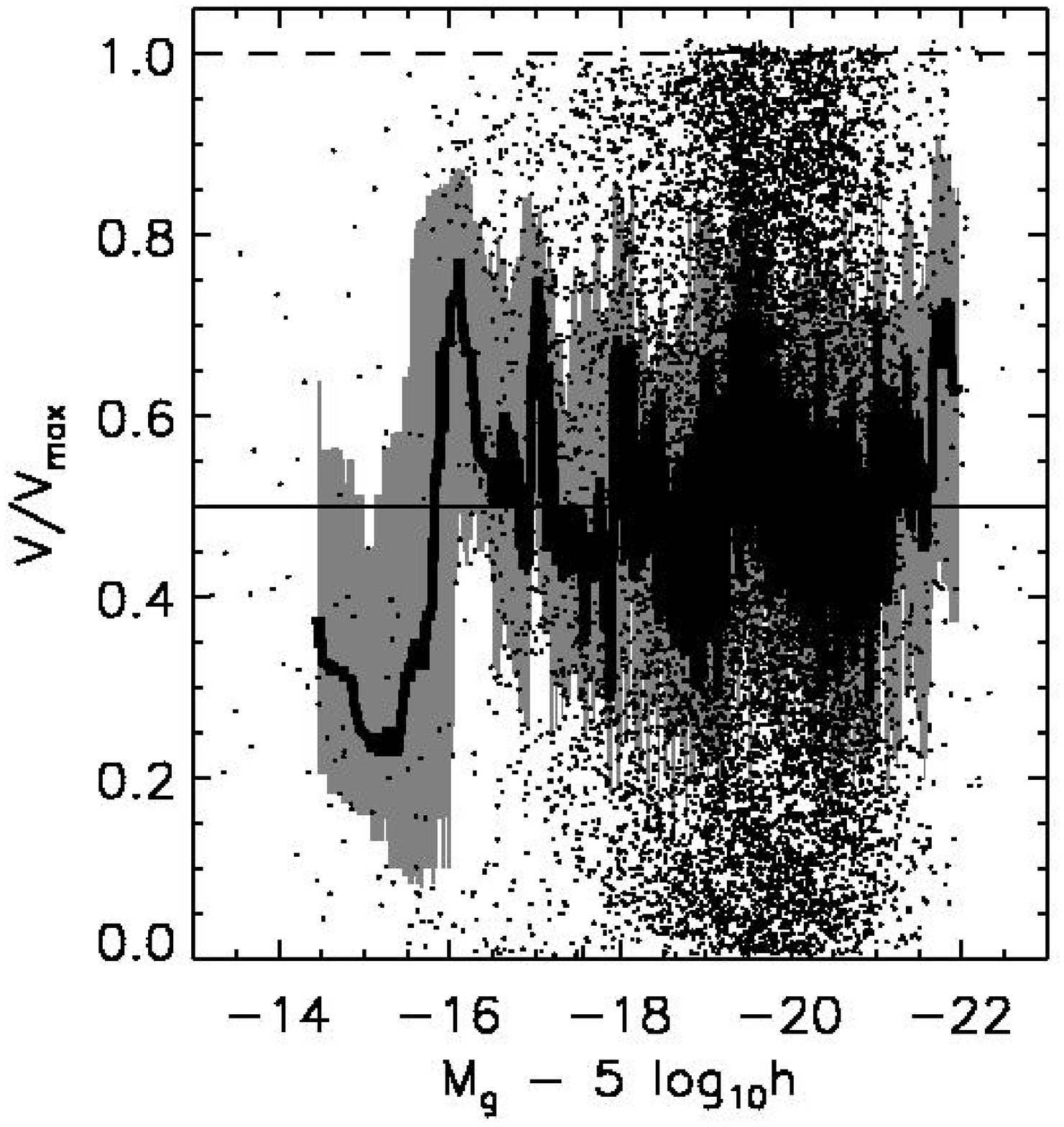}
\caption{\label{fig:compg} $V/V_{\rm max}$ against $g$-band 
absolute magnitude.  The median ({\it thick solid line}), and 
upper and lower quartiles ({\it shaded area}), are shown as a function
of $g$-band absolute magnitude.  The 
average value for the whole sample is $0.509\pm 0.003$, reasonably
consistent with the expected value of 0.5 ({\it thin solid line}).  }
\end{figure}

\begin{figure}[tb]
\vspace{-0.5cm}
\hspace{-0.5cm}
\epsfbox{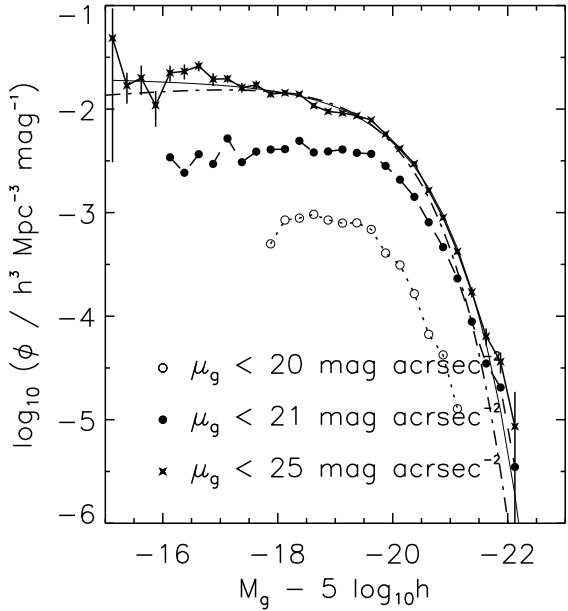}
\vspace{-0.2cm}
\caption{\label{fig:gbandtotal} $g$-band LFs using
different surface brightness cuts.
The solid line with data points represents
the total sample ($\mu_g < 25$\,mag\,arcsec$^{-2}$) LF.  The dotted and dashed
lines represent the LF for $\mu_g < 20$\,mag\,arcsec$^{-2}$ and
$\mu_g < 21$\,mag\,arcsec$^{-2}$ subsamples, which
shows the LF steepening at the faint end as the
surface brightness limit gets fainter. 
The dash-dot line shows the $g^{z=0.1}$ LF of \citet{blanton03}
transformed to redshift zero assuming unchanging $\phi^*$ and 
$\alpha$, and following their Table 10. 
The thin solid line is the Schechter fit
to our total LF, described in Table \ref{tab:fits}. }
\end{figure}

\begin{figure}[tb]
\vspace{-0.5cm}
\hspace{-0.5cm}
\epsfbox{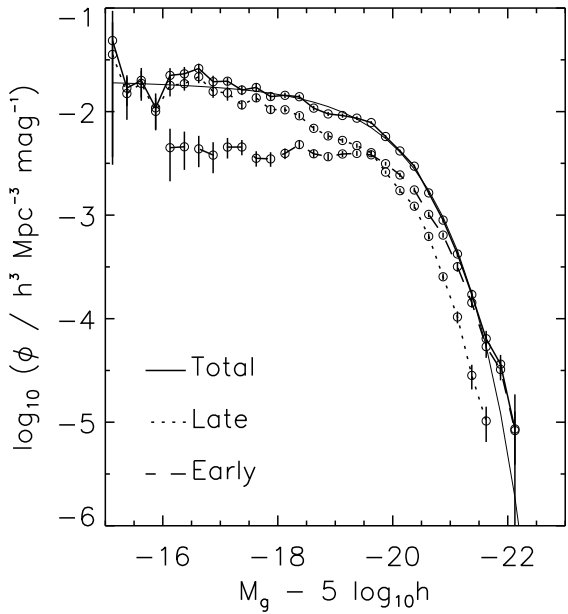}
\vspace{-0.2cm}
\caption{\label{fig:gbandsplit} $g$-band LF
split by morphological type.  
The solid line represents
the total LF.  The dotted and dashed
lines represent the LF for late and early-type
galaxies, separated using
$c_r = 2.6$.   }
\end{figure}

One strength of our combined SDSS and 2MASS sample is that
we can construct LFs in the optical
$ugriz$ passbands to accompany our NIR $K$-band LF.  
In this section, we derive a $g$-band 
limited LF.  The Schechter fits for other passbands are
given in Table \ref{tab:fits} for reference, and are discussed 
further in \S \ref{ld}.
Analogous to the $K$-band limited sample, we show the 
distribution of galaxy $V/V_{\rm max}$ with absolute
$g$-band magnitude (Fig.\ \ref{fig:compg}), the 
$g$-band LF derived using different surface brightness limits
(Fig.\ \ref{fig:gbandtotal}),
and the $g$-band LF split by morphological type (Fig.\ \ref{fig:gbandsplit}).

The $g$-band limited sample has 
$\langle V/V_{\rm max} \rangle = 0.509\pm0.003$,
which is slightly higher (by $\sim 3 \sigma$) than the purely random
distribution expectation of 0.5.  Nevertheless, as with the
$K$ sample, this departure
is small and should not affect our results at more than 
the few percent level.\footnote{We note that there is substantial 
structure in the $V/V_{\rm max}$ distribution at $g > -16$.  This
structure is likely caused by large scale structure, owing to the 
small $V_{\rm max}$ characteristic of galaxies with faint
absolute magnitudes in apparent magnitude-limited samples.
Furthermore, this structure is the probable origin of the fluctuations
in the $g$-band LF at $g > -16$ in Figs.\ \ref{fig:gbandtotal}
and \ref{fig:gbandsplit}.}
In Fig.\ \ref{fig:gbandtotal}, our $g$-band LF 
(solid line) compares well with that of \citet{blanton03},
shown as the dash-dotted line.  Furthermore, our $g$-band
luminosity density (see Table \ref{tab:fits}) is $\sim 7$\% larger than  
Blanton et al.'s value.  Blanton et al.\ did not include
light lost from the low surface brightness 
wings of early-type galaxies in their
luminosity density estimate, however.  When we account for 
the differences in technique by either neglecting 
the correction in our own analysis (resulting in a 
reduction of our luminosity density estimate by 4\%) or by comparing to 
Blanton et al.'s estimated correction (making Blanton et al.'s estimate
3\% higher), the agreement between our estimate and Blanton et al.'s
is well within the expected uncertainties.  
It is worth noting that, similar to the $K$-band LF, the
inclusion of lower surface brightness galaxies in $g$-band
increases $\phi^*$ slightly
and gives a steeper $\alpha$.

In Fig.\ \ref{fig:gbandsplit}, we show the $g$-band LF split
into early and late morphological types using $c_r=2.6$.
We find that the LF for early-types has a brighter $L^*$ and 
a flatter faint end slope than the later types, in 
agreement with many other studies of the local 
Universe \citep[e.g.,][]{bromley98,blanton01}.  We find 
a flat faint-end slope for the early-type LF,
which disagrees with the almost log-normal distribution 
seen for some local Universe early-type samples
\citep[e.g.,][]{blanton01,wolf03}.  
Some of this discrepancy is almost
certainly caused by a different sample selection.
For example, a cut at a relatively red constant
color will produce a log-normal LF because of the exclusion of faint,
genuinely old early-type galaxies that are too blue 
to satisfy the color cut owing to their low metallicity
\citep[e.g.,][]{ble92}.  
In the next section, we show that the early-type LF selected
from a magnitude-dependent color cut, which accounts for the CMR,
is also relatively flat \citep[see also][]{combo17}.

\subsection{Color-selection of early and late types} \label{sec:col}

\begin{figure*}[tb]
\center{\includegraphics[scale=1.0, angle=0]{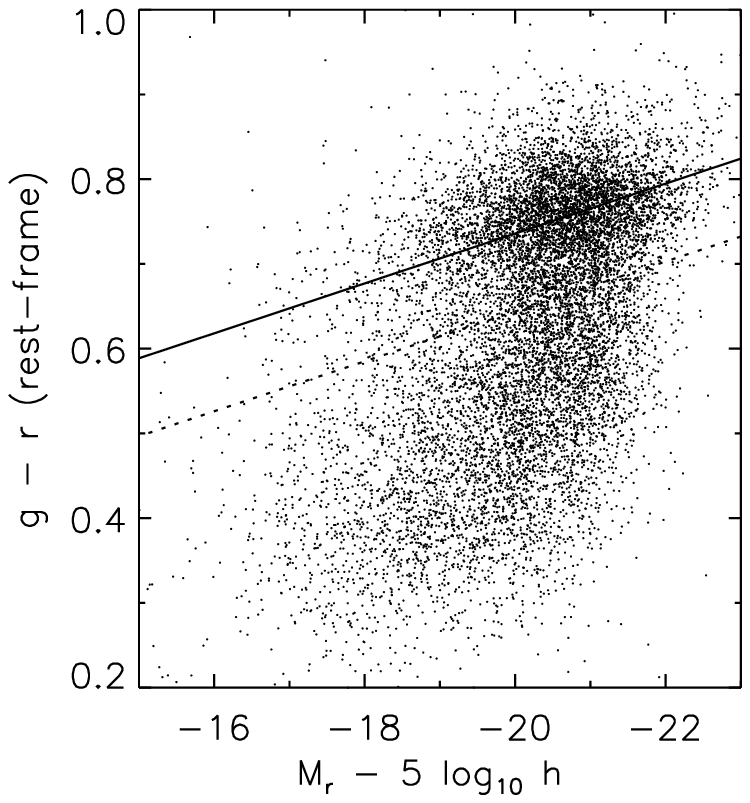}
\includegraphics[scale=1.0, angle=0]{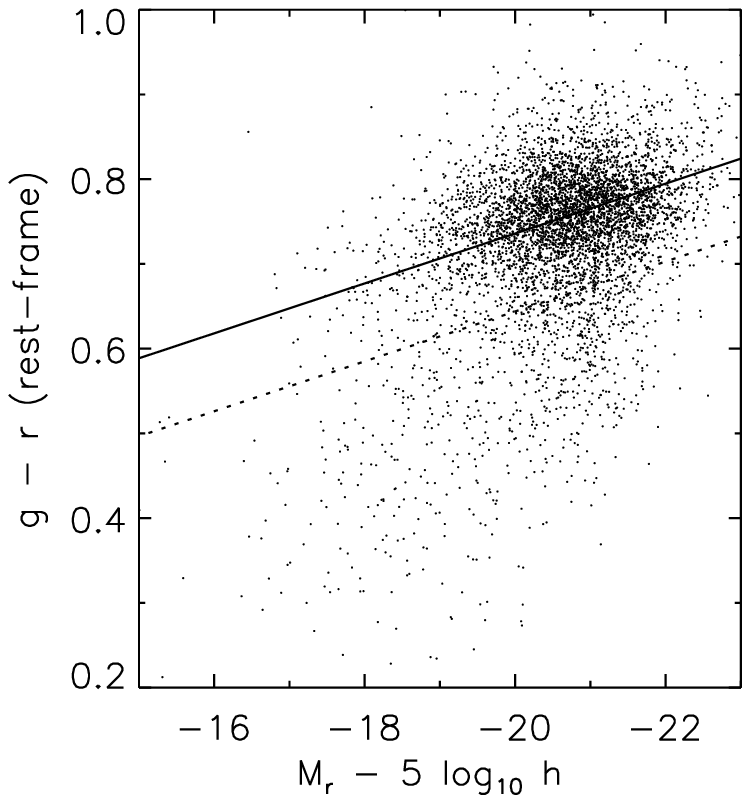}}
\caption{\label{fig:colormag} The CMR
of the $g$-band selected galaxies ({\it left}), and the
$g$-band selected subsample of early-types (with $c_r \ge 2.6$; {\it right}). 
Overplotted is 
the `fit' to the CMR with the same slope as local clusters ({\it solid line}),
and the color criterion for early-type selection ({\it dotted line}), which
corresponds to 0.092 mag bluer than the CMR fit.}
\end{figure*}

\begin{figure}[tb]
\vspace{-0.5cm}
\hspace{-0.5cm}
\epsfbox{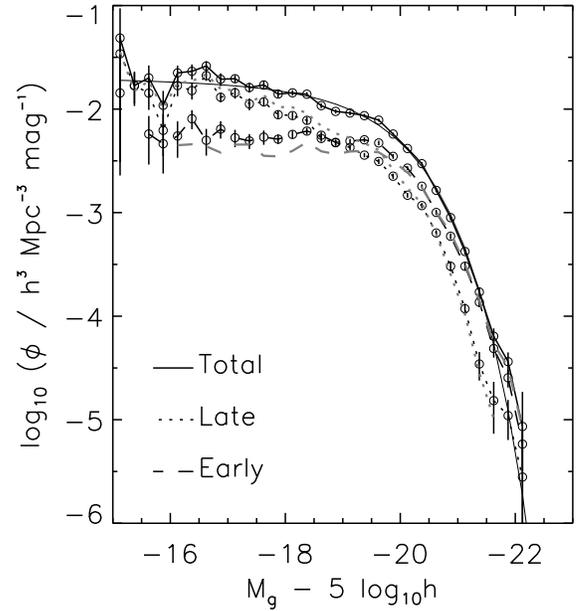}
\vspace{-0.2cm}
\caption{\label{fig:mfcol} 
$g$-band LF of color-selected galaxy types.
The solid line gives the total sample LF as in Fig.\ \ref{fig:gbandsplit}.
The black
dashed and dotted lines show the LF of color-selected early and 
late-type galaxies.  The corresponding grey lines show the 
$r$-band concentration parameter-selected samples from 
Fig.\ \ref{fig:gbandsplit}.  }
\end{figure}

We explore the role of morphological selection further
by using the broad-band colors of galaxies.  
We show the CMR of all
$g$-band selected galaxies in the left-hand panel
of Fig.\ \ref{fig:colormag}.  Here, one can clearly see
the `bimodality' of the color distribution 
of galaxies 
\citep[e.g.][]{strat,blanton03c,hogg03,combo17}.
The galaxies separate into coarse blue and red `sequences'.
The blue sequence 
has redder colors at brighter magnitudes, reflecting the older ages,
higher metallicities, and greater dust content in
brighter, more massive late-type galaxies
\citep[e.g.,][]{tully98,bdj}.  The red sequence
is also redder at brighter magnitudes, reflecting
a metallicity-magnitude relation for
older stellar populations \citep[e.g.,][]{ble92,ka97}.
The red sequence of galaxies from the field environment
is known to contain
predominantly early-type galaxies \citep{schweizer92,hogg03}.
Thus, choosing galaxies along the red sequence
is an excellent alternative method for selecting
early-type galaxies.

In Fig.\ \ref{fig:colormag}, the $g,r$ CMR
of early-type galaxies has been marked
using a solid line with a slope of $-0.03$. This $g-r$ slope is 
transformed from the CMR slope
in local galaxy clusters \citep{ble92,ka97}
of $-0.08$ in $U-V$ color using 
the \pegase stellar population synthesis models. 
This is in good agreement with the slope of $-0.02$ to $-0.04$ 
derived by \citet{bernardi_iv}.  We define
early-type galaxies as having colors redder than $\Delta (g-r) = -0.092$ 
mag from CMR ridge line,
i.e., everything above the dotted line in Fig.\ \ref{fig:colormag}, 
which corresponds to
$\Delta (U-V) = 0.25$ mag following \citet{combo17}.

We compare our color-magnitude based definition with the $c_r \geq 2.6$
subsample of early-type galaxies
in the right-hand panel of Fig.\ \ref{fig:colormag}.  Clearly
the majority of concentrated galaxies have colors that are
indicative of old stellar populations.  Fully 84\% of 
concentrated galaxies have colors that are redder than 
our color cut. Furthermore, the fraction 
of concentrated galaxies satisfying the color cut increases
towards brighter absolute magnitude, meaning that 
the overwhelming majority of the luminosity density in concentrated
galaxies will be from galaxies on the CMR.
Conversely,  70\% of the color-selected
early-types have $c_r\geq2.6$, although concentration has limitations
as a morphological-classifier (\S \ref{sec:cnc}).

We show the $g$-band LF of early and late-types defined using our
color cut in Fig.\ \ref{fig:mfcol}.
Although the color selection gives a larger number of
early-type systems, the overall differences between early and
late-type LFs are similar to those we find using concentration to divide
our galaxy sample morphologically (\S \ref{gband:samp}).
Thus, the basic result
that late-types have fainter $L^*$ and steeper
faint-end slopes than early-types is robust to
different type definitions \citep[see also][]{kochanek01}.  

\subsection{Comparing luminosity density estimates of the local Universe} \label{ld}

In Table \ref{tab:fits}, we see that the luminosity densities we find
for the $ugri$ passbands agree well
with the $z=0$ densities of \citet{blanton03}.
They compare their LFs
and luminosity densities in detail with various other
local Universe determinations, finding agreement
to within 10\% (whether or not one corrects
for the extrapolation of early-type galaxy magnitudes to total
in SDSS).  In the $z$-band, we find a $\sim 15$\% 
lower luminosity density than \citet{blanton03}.
We agree well with Blanton et al.'s estimate at $z = 0.1$ of
$2.69 \pm 0.05 \times 10^8\,h\,L_{\sun}$\,Mpc$^{-3}$; therefore,
the origin of the $z$-band discrepancy at $z\sim0$ is likely 
an unphysically large
density evolution in their analysis.
Additionally, our K-band luminosity density agrees with those
of \citet{cole01} and \citet{kochanek01}.
They have also compared
their LFs with literature determinations and find typically excellent
agreement.  

\citet{wright01} claims an overall discrepancy between 
the optical LF determinations of \citet{blanton01} and 
the NIR LFs of \citet{cole01} or \citet{kochanek01}, finding 
a factor of $\sim 2$ more luminosity
density in the optical than in the NIR.  Indeed, \citet{huang03} 
claim that they `solve' this problem by declaring that the 
the local $K$-band luminosity density is a factor of 
two higher.  However, \citet{blanton03} show that 
the luminosity density estimates of \citet{blanton01} were
systematically too high by $\ga 60$\% (more in the bluer passbands).  
We investigate this issue through a luminosity density
comparison using our galaxy sample in different passbands.

\begin{figure}[tb]
\vspace{-0.5cm}
\hspace{-0.5cm}
\epsfbox{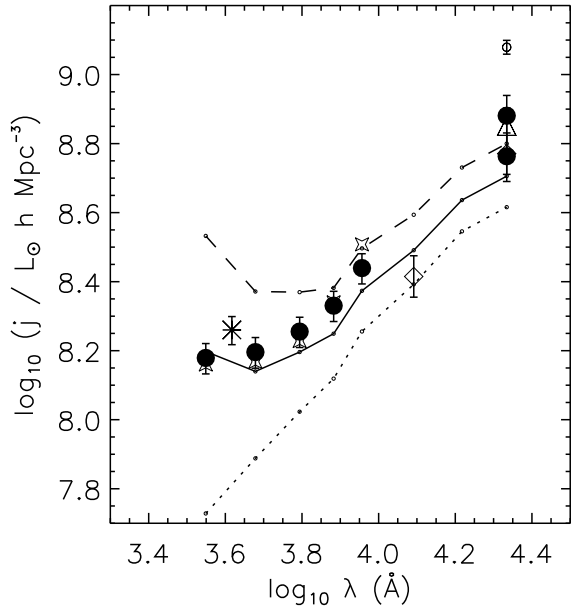}
\vspace{-0.2cm}
\caption{\label{fig:lumdens} 
The luminosity density of galaxies in the local Universe. 
Solid symbols show the results from our LF 
determinations in the $ugrizK$ passbands. Two estimates are presented 
in $K$-band --- our direct estimate ({\it lower solid point}) 
and our estimate which accounts for 2MASS's bias against
LSB galaxies ({\it upper solid point}).  We also plot
a number of literature determinations of 
local luminosity density: $ugriz$ from \citet[{\it stars}, 
which for $u$, $g$ and 
$r$ are under the solid circles]{blanton03}; $b_J$ from 
\citet[{\it asterisk}]{norberg02}; $J$ and $K$-bands from 
\citet[{\it diamonds}]{cole01}; $K$-band from 
\citet[{\it triangle}]{kochanek01};
and the $K$-band estimate of \citet[][{\it small circle}]{huang03}.
We argue that the \citet{huang03} result is artificially high
(see text).  We overplot
three stellar population models for 12 Gyr old stellar
populations of solar metallicity, formed with exponentially-decreasing 
SFRs with $e$-folding times $\tau$ of 2 Gyr ({\it dotted
line}), 4 Gyr ({\it solid line}), and 8 Gyr ({\it dashed line}). }
\end{figure}

We show the luminosity density of galaxies in the local Universe
in Fig.\ \ref{fig:lumdens}.  The formal uncertainties are 
added in quadrature to a density and magnitude systematic uncertainty
of 10\% 
for the optical data, and 15\% 
for the $K$-band point (because of the additional
10\% error in extrapolating to total; Table \ref{tab:sys}).
The solid symbols show the results of our analysis, and the open 
symbols a variety of literature determinations of luminosity
density, including the 
$K$-band determination of \citet{huang03}, which 
we have argued is artificially high.  
Overplotted are three stellar population models for 12 Gyr old stellar
populations of solar metallicity, formed with exponentially-decreasing 
SFRs with $e$-folding times $\tau$ of 2 Gyr ({\it dotted
line}), 4 Gyr ({\it solid line}) and 8 Gyr ({\it dashed line}).
All of the models are constrained to have the same stellar mass density 
as we derive in \S \ref{sec:mf}.
Clearly our data reproduce many of the 
luminosity density determinations in the local Universe
between $u$ and $K$-bands, but with the dual advantages that we use
one consistent dataset to determine the luminosity
density in all passbands, and that we understand
sources of bias in 2MASS better than previous work owing to
the multi-passband coverage offered by SDSS and 2MASS.  
Furthermore, the shape of the cosmic mean spectrum is broadly consistent
with a relatively metal-rich galaxy with a SFH peaked
at early times and decreasing to a present day non-zero rate.
This is in excellent agreement with the
work of \citet{baldry02} and \citet{glazebrook03}.

Fig.\ \ref{fig:lumdens} illustrates that we have a good
understanding of the luminosity density of 
the local Universe.  Thus, it is worth spending a few words on why
there has been such confusion surrounding its determination.  
\citet{blanton01} presented
luminosity densities around a factor of two higher
than the new, revised SDSS estimates; this discrepancy is caused primarily by 
evolution correction uncertainties
\citep{blanton03}.  These high estimates are at variance
with earlier luminosity density determinations at optical
\citep[see, e.g., Fig. 1 of][]{wright01} and NIR wavelengths
\citep[e.g.,][]{gardner97,kochanek01}.
We now see that when evolution,
$k$-corrections, and systematic bias are properly and carefully accounted for 
\citep[as in this work, or][]{blanton03},
the luminosity density of the local Universe
between 3500{\AA} and 2.2{\micron} 
is understood to within $\sim20\%$ at a given 
wavelength.  

\section{Stellar Mass Functions} \label{sec:mf}

\begin{figure}[tb]
\vspace{-0.5cm}
\hspace{-0.5cm}
\epsfbox{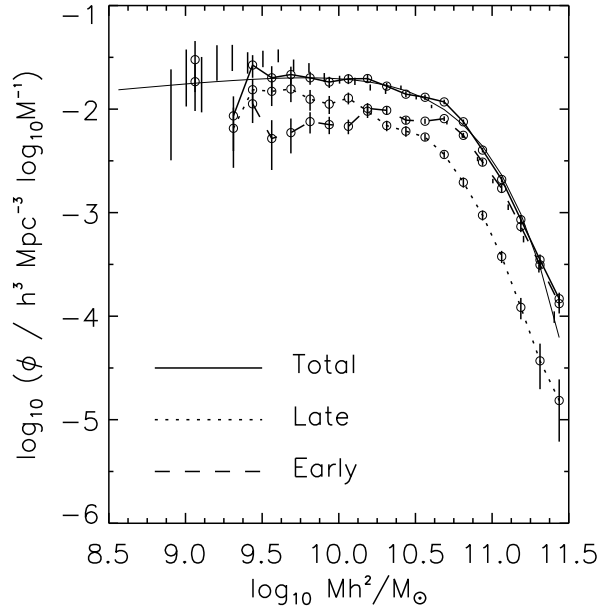}
\vspace{-0.2cm}
\caption{\label{fig:kmasssplit} $K$-band derived 
stellar MF.
The solid line represents
the total MF.  The dotted and dashed
lines represent the MF for late and early-type
galaxies, separated using the
$c_r = 2.6$ criteria.  The naked error bars
denote the 2MASS based MF of 
\protect\citet{cole01}, corrected to our IMF.
The thin solid line is our Schechter function fit
to the MF.}
\end{figure}

\begin{figure}[tb]
\vspace{-0.5cm}
\hspace{-0.5cm}
\epsfbox{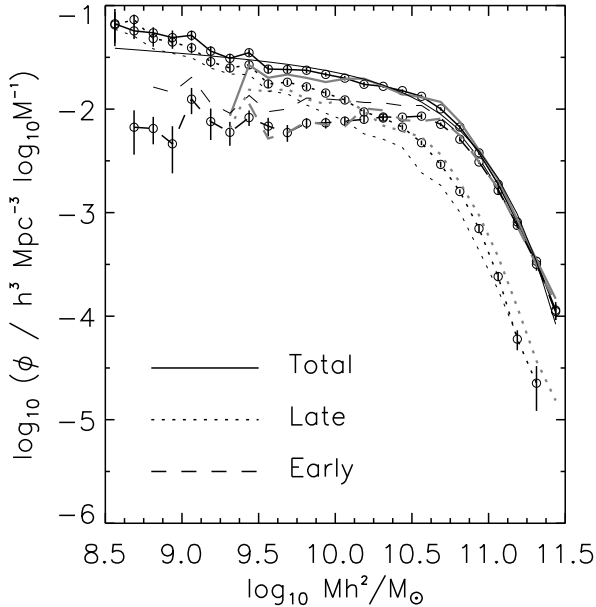}
\vspace{-0.2cm}
\caption{\label{fig:gmasssplit} $g$-band derived 
stellar MF.
The solid line represents
the total MF.  The black dotted and dashed
lines represent the MF for late and
early-type galaxies, separated using the
$c_r = 2.6$ criteria.  The thin solid line is our Schechter function fit
to the MF.  Overplotted in grey are the $K$-band derived 
stellar MFs for the total sample and the two morphological subsamples
from Fig.\ \ref{fig:kmasssplit}.
The thin black dashed and dotted lines show the $g$-band MFs of
color-selected early and late-type galaxies.  The data points
included in this plot are tabulated
in Table \ref{tab:gmf}.}
\end{figure}

\begin{table*}
\caption{Galaxy Stellar Mass Function Fits {\label{tab:mffits}}}
\begin{center}
\begin{tabular}{lcccc}
\hline
\hline
Passband & 
$\phi^*$ & $\log_{10} M^*\,h^2$ &
$\alpha$ & $\rho$  \\
 &  $h^3$\,Mpc$^{-3}\,\log_{10}M^{-1}$ & $M_{\sun}$ & 
& $h\,M_{\sun}$\,Mpc$^{-3}$  \\
\hline
\multicolumn{5}{c}{All Galaxies} \\
\hline
$g$ & 0.0102(5) & 10.70(2) & $-$1.10(2) & 
5.47(11)$\times 10^8$  \\
$K$ & 0.0133(6) & 10.63(1) & $-$0.86(4) 
&  5.26(12)$\times 10^8$ \\
\hline
\multicolumn{5}{c}{Early-Type Galaxies}\\
\hline
$g$ & 0.0083(4) & 10.62(2) & $-$0.60(4) & 
3.08(6)$\times 10^8$  \\
$g_{\rm col}$ & 0.0107(8) & 10.60(4) & $-$0.70(7) & 
3.84(9)$\times 10^8$  \\
$K$ & 0.0089(4) & 10.61(2) & $-$0.52(7) 
& 3.19(8)$\times10^8$  \\
\hline
\multicolumn{5}{c}{Late-Type Galaxies}\\
\hline
$g$ & 0.0059(3) & 10.51(2) & $-$1.27(3) & 
2.40(4)$\times 10^8$  \\
$g_{\rm col}$ & 0.0027(2) & 10.59(3) & $-$1.45(3) & 
1.70(4)$\times 10^8$  \\
$K$ & 0.0071(6) & 10.48(3) & $-$0.94(8) 
& 2.10(9)$\times10^8$  \\
\hline
\end{tabular}
\\ 
\vspace{-1.0cm}
\tablecomments{A `diet' Salpeter IMF is used for the above MFs,
which satisfies maximum disk constraints from galaxies in the 
Ursa Major galaxy cluster \citep{ml}.
For reference, the Stellar mass density of \citet{cole01}, 
corrected to our IMF and for
light missed 
by the 2MASS aperture, is $\sim$5.5$\times10^8 h M_{\sun}$\,Mpc$^{-3}$.
Formal error estimates are in parentheses.
These MFs are systematically uncertain at the 30\% level (in terms of mass)
due to uncertainties in density normalization, absolute magnitude calibration,
SFHs, and the role of dust (\S \ref{sec:sysml}).
To convert these MFs to different IMFs, add a constant
to the $\log_{10} M^*$ following \S \ref{sec:imf}.  See Table \ref{tab:sys}
for further discussion of the systematic errors.  }
\end{center}
\end{table*}

\begin{table*}
\caption{$g$-band-Selected Galaxy Stellar Mass Functions {\label{tab:gmf}}}
\begin{center}
\begin{tabular}{lccccc}
\hline
$\log_{10} Mh^2/M_{\sun}$ & \multicolumn{5}{c}{$\log_{10} 
    (\phi/h^3\,{\rm Mpc^3}\,\log_{10}M^{-1})$} \\
& Total & Early ($c_r \ge 2.6$) & Late ($c_r < 2.6$) & Early (color) &
        Late (color) \\
\hline
\hline
 8.56 & $-1.176_{-0.240}^{+ 0.154}$ & \nd & $-1.187_{-0.173}^{+ 0.123}$ & \nd & $-1.229_{-0.147}^{+ 0.109}$ \\ 
 8.69 & $-1.245_{-0.078}^{+ 0.066}$ & $-2.175_{-0.265}^{+ 0.163}$ & $-1.135_{-0.054}^{+ 0.048}$ & \nd & $-1.296_{-0.079}^{+ 0.067}$ \\ 
 8.81 & $-1.264_{-0.071}^{+ 0.061}$ & $-2.187_{-0.153}^{+ 0.113}$ & $-1.319_{-0.079}^{+ 0.067}$ & $-1.787_{-0.165}^{+ 0.119}$ & $-1.418_{-0.067}^{+ 0.058}$ \\ 
 8.94 & $-1.309_{-0.089}^{+ 0.074}$ & $-2.336_{-0.285}^{+ 0.170}$ & $-1.352_{-0.065}^{+ 0.057}$ & $-1.839_{-0.274}^{+ 0.167}$ & $-1.461_{-0.053}^{+ 0.048}$ \\ 
 9.06 & $-1.288_{-0.049}^{+ 0.044}$ & $-1.904_{-0.145}^{+ 0.109}$ & $-1.408_{-0.054}^{+ 0.048}$ & $-1.693_{-0.172}^{+ 0.123}$ & $-1.505_{-0.033}^{+ 0.031}$ \\ 
 9.19 & $-1.440_{-0.041}^{+ 0.037}$ & $-2.120_{-0.178}^{+ 0.126}$ & $-1.542_{-0.054}^{+ 0.048}$ & $-1.942_{-0.130}^{+ 0.100}$ & $-1.604_{-0.049}^{+ 0.044}$ \\ 
 9.31 & $-1.510_{-0.031}^{+ 0.029}$ & $-2.224_{-0.131}^{+ 0.100}$ & $-1.603_{-0.043}^{+ 0.039}$ & $-2.040_{-0.103}^{+ 0.083}$ & $-1.662_{-0.038}^{+ 0.035}$ \\ 
 9.44 & $-1.455_{-0.024}^{+ 0.022}$ & $-2.082_{-0.081}^{+ 0.068}$ & $-1.572_{-0.039}^{+ 0.036}$ & $-1.868_{-0.081}^{+ 0.068}$ & $-1.667_{-0.039}^{+ 0.036}$ \\ 
 9.56 & $-1.614_{-0.033}^{+ 0.030}$ & $-2.164_{-0.098}^{+ 0.080}$ & $-1.757_{-0.040}^{+ 0.037}$ & $-2.025_{-0.095}^{+ 0.078}$ & $-1.827_{-0.037}^{+ 0.034}$ \\ 
 9.69 & $-1.618_{-0.026}^{+ 0.024}$ & $-2.227_{-0.089}^{+ 0.074}$ & $-1.740_{-0.027}^{+ 0.026}$ & $-1.999_{-0.061}^{+ 0.054}$ & $-1.851_{-0.035}^{+ 0.033}$ \\ 
 9.81 & $-1.625_{-0.021}^{+ 0.020}$ & $-2.135_{-0.055}^{+ 0.049}$ & $-1.785_{-0.032}^{+ 0.029}$ & $-1.894_{-0.051}^{+ 0.045}$ & $-1.960_{-0.038}^{+ 0.035}$ \\ 
 9.94 & $-1.665_{-0.028}^{+ 0.026}$ & $-2.133_{-0.055}^{+ 0.049}$ & $-1.845_{-0.028}^{+ 0.027}$ & $-1.902_{-0.034}^{+ 0.032}$ & $-2.040_{-0.032}^{+ 0.030}$ \\ 
10.06 & $-1.701_{-0.024}^{+ 0.022}$ & $-2.116_{-0.047}^{+ 0.042}$ & $-1.912_{-0.032}^{+ 0.030}$ & $-1.915_{-0.033}^{+ 0.031}$ & $-2.111_{-0.029}^{+ 0.027}$ \\ 
10.19 & $-1.761_{-0.018}^{+ 0.017}$ & $-2.099_{-0.033}^{+ 0.031}$ & $-2.027_{-0.028}^{+ 0.026}$ & $-1.936_{-0.037}^{+ 0.034}$ & $-2.239_{-0.027}^{+ 0.026}$ \\ 
10.31 & $-1.780_{-0.026}^{+ 0.025}$ & $-2.076_{-0.032}^{+ 0.029}$ & $-2.086_{-0.027}^{+ 0.026}$ & $-1.940_{-0.025}^{+ 0.024}$ & $-2.291_{-0.025}^{+ 0.023}$ \\ 
10.44 & $-1.822_{-0.021}^{+ 0.020}$ & $-2.078_{-0.020}^{+ 0.019}$ & $-2.174_{-0.026}^{+ 0.025}$ & $-1.967_{-0.025}^{+ 0.023}$ & $-2.370_{-0.022}^{+ 0.020}$ \\ 
10.56 & $-1.877_{-0.018}^{+ 0.017}$ & $-2.068_{-0.028}^{+ 0.026}$ & $-2.325_{-0.022}^{+ 0.021}$ & $-1.964_{-0.019}^{+ 0.018}$ & $-2.614_{-0.023}^{+ 0.021}$ \\ 
10.69 & $-1.998_{-0.019}^{+ 0.018}$ & $-2.147_{-0.018}^{+ 0.017}$ & $-2.534_{-0.029}^{+ 0.027}$ & $-2.086_{-0.016}^{+ 0.015}$ & $-2.736_{-0.033}^{+ 0.031}$ \\ 
10.81 & $-2.173_{-0.020}^{+ 0.019}$ & $-2.291_{-0.022}^{+ 0.021}$ & $-2.796_{-0.021}^{+ 0.020}$ & $-2.244_{-0.017}^{+ 0.016}$ & $-2.995_{-0.051}^{+ 0.046}$ \\ 
10.94 & $-2.422_{-0.021}^{+ 0.020}$ & $-2.511_{-0.018}^{+ 0.017}$ & $-3.154_{-0.039}^{+ 0.036}$ & $-2.475_{-0.019}^{+ 0.018}$ & $-3.366_{-0.049}^{+ 0.044}$ \\ 
11.06 & $-2.726_{-0.020}^{+ 0.019}$ & $-2.785_{-0.034}^{+ 0.031}$ & $-3.617_{-0.045}^{+ 0.041}$ & $-2.769_{-0.023}^{+ 0.022}$ & $-3.747_{-0.057}^{+ 0.050}$ \\ 
11.19 & $-3.090_{-0.040}^{+ 0.036}$ & $-3.124_{-0.033}^{+ 0.031}$ & $-4.222_{-0.107}^{+ 0.086}$ & $-3.126_{-0.037}^{+ 0.034}$ & $-4.192_{-0.164}^{+ 0.119}$ \\ 
11.31 & $-3.470_{-0.050}^{+ 0.045}$ & $-3.500_{-0.062}^{+ 0.054}$ & $-4.647_{-0.268}^{+ 0.165}$ & $-3.482_{-0.052}^{+ 0.047}$ & \nd \\ 
11.44 & $-3.942_{-0.099}^{+ 0.081}$ & $-3.955_{-0.082}^{+ 0.069}$ & \nd & $-3.967_{-0.136}^{+ 0.103}$ & $-5.180_{-0.967}^{+ 0.335}$ \\ 
\hline
\end{tabular}
\\ 
\vspace{-1.0cm}
\tablecomments{These MFs are systematically uncertain
at the 30\% level in terms of mass
due to uncertainties in the raw LFs, SFHs, 
and the role of dust (\S \ref{sec:sysml}). 
To convert these MFs from our default `diet' Salpeter IMF to a 
different stellar IMF, add a constant 
to the $\log_{10} M$ following \S \ref{sec:imf}.}

\end{center}
\end{table*}

One of the strengths of our combined SDSS and 2MASS sample
is that there are accurately measured colors for all the 
sample galaxies.  Under the assumption of a universally-applicable stellar IMF,
we can then constrain the stellar M/L ratios to within
25\% in a systematic sense, given the uncertainties in galaxy
age, dust content, and the role of bursts of SF (\S \ref{sec:sysml}). 
From stellar M/L ratios, we
construct galaxy stellar MFs
from both the $K$-band limited and optically-limited 
galaxy samples, which allows us to explore potential sources of 
systematic bias caused by the choice of passband.

In Fig.\ \ref{fig:kmasssplit}, we show an estimate of the stellar MF from 
our $K$-band 
limited sample.  The solid line
shows the stellar MF for all galaxies.  The dashed and dotted lines show
early and late-type galaxies, respectively.  A thin solid line
denotes the Schechter function fit to the 
stellar MF, and the naked error bars show the $K$-band-derived stellar
MF from \citet{cole01}.  When expressed in terms of stellar mass, the
early-type galaxies have a higher characteristic mass
$M^*$, and a shallower faint end slope $\alpha$, than the later
types.  Furthermore,
the stellar MF that we derive is
in excellent agreement with the estimate of \citet{cole01}.
Integrating under the MF, we find that
our total $K$-band stellar mass density estimate is 
$5.3 \pm 0.1 \pm 1.6 \times 10^8\,h\,M_{\sun}\,{\rm Mpc}^{-3}$ 
(random and systematic errors, respectively; see \S \ref{sec:sysml}), 
in excellent agreement with the $\sim 5.5\pm0.8 \times 
10^8\,h\,M_{\sun}\,{\rm Mpc}^{-3}$
estimate of \citet{cole01}.  

To explore 2MASS's bias against LSB galaxies in a complementary way,
we show a $g$-band derived stellar MF in 
Fig.\ \ref{fig:gmasssplit}.  The solid, dashed, and dotted 
black lines show the $g$-band stellar MF for all galaxies, 
early-types and late-types, respectively.  The grey
lines show the results from Fig.\ \ref{fig:kmasssplit}
for the $K$-band limited sample.  
A cursory inspection shows that 
the $g$-band MF is relatively poorly fit by a Schechter
function; the break in the MF is too sharp, and the 
stellar MF faintwards of $Mh^2 \sim 3 \times 10^{10}M_{\sun}$
is better fit by a single power law than a Schechter function.
For this reason, we present the $V/V_{\rm max}$ 
data points for the $g$-band derived stellar mass functions
in Table \ref{tab:gmf}.
The $g$-band stellar MF shows excellent agreement
at the knee of the MF with the $K$-band stellar MF, and 
shows a steeper faint-end slope (below $10^{10}h^2 M_{\sun}$),
in agreement with our earlier prediction of the `real'
$K$-band LF.  This again argues that 2MASS misses 
faint LSB galaxies (\S \ref{samp:compl}).
The $g$-band stellar MF continues to much lower
masses with better signal-to-noise than the $K$-band stellar MF, showing
that the stellar MF has a relatively steep faint end ($\alpha \la -1.1$).
A steep stellar MF contrasts with most contemporary
determinations of galaxy LFs over cosmologically-significant
volumes, which have $\alpha>-1.1$ 
\citep[e.g.,][]{cole01,norberg02,blanton03}.
We expect a steeper stellar MF because late-type galaxies
have lower stellar M/L ratios than earlier types, and there appears to be an
increasing contribution from later types at low luminosities.  
An interesting implication of the Universal stellar MF is
that, because of the strong variation in optical 
stellar M/L ratios with galaxy stellar mass \citep[e.g.,][]{kauffmann03b},
the faint end slope of optical LFs should be {\it shallower} than
the faint end slope of NIR LFs.  Deeper optical and especially NIR LFs
will be in a good position to explore this issue in more detail.

\vspace{1cm}

\section{Discussion} \label{sec:disc}

\subsection{Changing the Stellar IMF} \label{sec:imf}

\citet{ml} find that the dominant source of uncertainty in 
estimating stellar M/L ratios from galaxy luminosities and optical
colors is the stellar IMF.  Uncertainties in the slope of 
the stellar IMF for high-mass stars leave
the color--M/L ratio correlation essentially unaffected.  In contrast, uncertainties
in the behavior of the stellar IMF for low-mass stars
affects the overall M/L ratio scale.  For example,
an IMF richer in low-mass stars yields a higher stellar M/L ratio at a given
color.  Note that the uncertainties in the low-mass end 
of the IMF do not affect
galaxy colors or luminosities, and hence, the color--M/L ratio correlation,
because these stars are too faint.

Our stellar
MFs are accurate to 30\%, if we understand how the 
stellar IMF behaves at low stellar mass.  This accuracy
is useful for relative comparisons such as our estimate versus
the stellar mass density of the Universe derived by 
\citet{fuku98} or \citet{glazebrook03}.
However, to determine in an absolute sense the stellar MF, or 
stellar mass density
of the Universe, or to compare with 
the Universal cold gas density, we must account for the full
range of stellar M/L ratio uncertainty from uncertainties in stellar IMF.

We want to quantify a reasonable range of stellar IMF uncertainty.  \citet{ml}
place a constraint on the stellar M/L ratios by demanding that the
stellar mass in the central parts of spiral galaxies in the 
Ursa Major cluster not over-predict their rotation velocities.
This `maximum-disk' constraint forces there to be less low-mass
stars than a \citet{salp} IMF, and motivates the 
`diet' Salpeter IMF. 
Given the {\it HST} key project distance scale \citep{freedman01},
the Salpeter IMF is too rich in low-mass stars to satisfy dynamical
constraints \citep[e.g.,][]{weiner01,kauffmann03a,kranz03}.
This result is strengthened by our earlier
comparison with estimates of the {\it total} M/L ratios
of early-types by \citet{bernardi_iii}, who find
a color--M/L ratio correlation consistent with our 
`diet' Salpeter IMF.  There are considerable
uncertainties in this kind of analysis, 
including aperture bias, the effects of a non-isotropic velocity
ellipsoid on the measured velocity dispersion 
\citep[e.g.,][]{cretton00}; nevertheless,
it is encouraging that maximum disks and `maximum spheroids' yield
consistent results to the best of our knowledge.
We therefore adopt
the `diet' Salpeter IMF as our default IMF, and the M/L ratios we derive
from colors, assuming this IMF, will be good upper limits to 
the stellar M/L ratio.  

Of course, the stellar M/L ratio can be lower than this maximal
value.  For example, \citet{bottema93,bottema97,bottema99} 
argue for a substantially sub-maximal M/L ratio for {\it all} disk-dominated
galaxies based on an analysis of the vertical velocity dispersion of
stars.  In a similar vein, \citet{courteau99} argue that all disks are 
sub-maximal,
based on a lack of surface-brightness dependence in the luminosity-linewidth
relation \citep{tully77}.  
In contrast, \citet{ath87}, \citet{weiner01},
and \citet{kranz03}, use a variety of techniques to
propose a scenario in which low rotation velocity 
galaxies are substantially sub-maximal \citep[in agreement 
with][]{bottema97,courteau99}, 
and high rotation velocity ($\sim$ massive)
galaxies are essentially maximum-disk.  The latter work
is quite consistent with a constant, maximum-disk constrained
IMF.  Thus, it is the influence of a dark matter halo that will give
sub-maximal M/L ratios for dark-matter dominated,
LSB, slower rotators 
\citep[e.g.,][]{deblok98}.\footnote{See Fig.\ 6 of \citet{ml} 
for an illustration of the higher maximum
disk M/L ratio estimates for LSB galaxies, 
plausibly indicating dark matter domination, even in the inner
parts of the galaxies.}  We defer a clearly merited 
more detailed discussion of this issue to a future paper
(de Jong \& Bell, in preparation).

Attacking the problem from another angle, there is 
considerable and comparable uncertainty in 
the determinations of the IMF slope for
low-mass stars in the Milky Way.  These uncertainties
are well-discussed in reviews by 
\citet{scalo98} and \citet{kroupa02}.  A fair assumption is
that a universal IMF exists
\citep[although see, e.g.,][for a differing view]{scalo98},
and the range of proposed IMFs, from \citet{salp} to 
\citet{kroupa93} to the 63\% of maximum-disk velocity 
\citep{bottema97}, should bracket the possible IMF
choices.  To convert our maximum-disk constrained
`diet' Salpeter IMF to \citet{salp}, \citet{gould97}, \citet{scalo86},
\citet{kroupa93}, \citet{kroupa02}, \citet{kennicutt83}, or
Bottema 63\% maximal IMFs, we should 
add roughly (0.15, 0.0, $-0.1$, $-$0.15, 
$-$0.15, $-$0.15, $-$0.35) dex to the stellar
M/L ratios predicted using the maximal IMF\footnote{Note that
the Salpeter case violates the maximum disk constraints, however.}.
Thus, including the full range of systematic
uncertainty, our stellar masses can be increased by $\sim 0.1$ dex, 
and decreased by $\sim 0.45$ dex, {\it in a systematic sense}.  
For comparison with other estimates of stellar mass density, 
our results are better constrained; they should be changed
to the same IMF by adjusting the zero-point, and the full range of systematic
uncertainty will then be $\pm 0.1$ dex.

\subsection{The Stellar Mass Density of the Universe}

It is interesting to consider at this stage the stellar mass density
of the Universe.  Adopting the `maximal' diet Salpeter IMF, 
we find that $\Omega_* h \sim 0.00197(4)$ (formal error), 
with a $\pm30$\% systematic error 
due to LF uncertainties and the effects of dust and bursts of SF.
This estimate of $\Omega_* h \sim 0.0020\pm0.0006$ is an upper limit;
if all galaxies are substantially sub-maximal, this estimate will 
need to be revised downwards.
Splitting these into early and late types (by either concentration
or color-selection), we find that 
$\Omega_{\rm *,Early} h \sim 0.0012(4)$ and 
$\Omega_{\rm *,Late} h \sim 0.0007(2)$.  This estimate accounts for 
classification and systematic stellar M/L ratio uncertainties, while
assuming the same stellar IMF in early and late types.
Thus, we find that between
half and three quarters of the stellar mass in the local Universe is in 
early types (this is robust to exactly {\it which} universal 
IMF is chosen, as long as it applies to both early and late types).  
Of course, much of the stellar mass in late-type galaxies
will also be old.  
This agrees with
the conclusions of \citet{hogg02}, who find a very similar
result, but accounting for stellar M/L ratios in a less elaborate way.
This conclusion is in {\it qualitative} agreement with the shape
of the `cosmic SFH' from direct 
\citep[e.g.][]{madau96,yan99,blain99}, or
indirect \citep{baldry02,glazebrook03}, estimates; all
estimates agree that the bulk of SF in the Universe
happened at early times, and is much slower at the present day.

Adopting the same `diet' Salpeter maximum-disk tuned IMF
as we do, \citet{cole01} find $\Omega_* h = 2.0\pm0.3 \times 10^{-3}$
accounting for light missed by 2MASS,  \citet{persic92} find
$\Omega_{\rm spirals+ellipticals} \sim 2\times 10^{-3}$, \citet{fuku96}
find $\Omega_* h \sim 2.6\pm1.3\times 10^{-3}$, \citet{kochanek01} 
find $\Omega_* h = 2.4\pm0.3\times 10^{-3}$, and \citet{glazebrook03}
find $\Omega_* h$ values between $1.8\times 10^{-3}$ and 
$3.9\times 10^{-3}$.  Thus, our determination of $2.0\pm0.6\times 10^{-3}$
is in excellent agreement with the literature determinations, and we
have the advantage that we have estimated our stellar M/L ratios
robustly, and our systematic uncertainties
are better understood.  We choose not to explore
the quantitative consistency between our results and the integrated
cosmic SFH, in part because the cosmic SFH is 
still poorly constrained in the particularly
important epoch $0.5 < z < 2$, where much of the stellar 
content in the Universe appears to have formed
\citep[e.g.,][]{madau96,blain99,haarsma00}.

\citet{bary} discuss implications of this result for the 
baryonic MF of galaxies, the cosmic mean density of 
stellar and cold gas in the local Universe, and the mean
`cold gas fraction' in the local Universe, accounting
again for the main sources of systematic uncertainty.  \citet{salucci}
present a complementary analysis of spiral
galaxies using primarily dynamically-derived M/L ratios, finding
a similar but slightly higher cosmic mean density 
of stellar and cold gas mass compared to \citet{bary}.  
Either way, both works show the exciting potential offered
by a detailed understanding of the mass-to-light
ratios and stellar MFs in the local Universe.

It is interesting to use the stellar mass density estimate
in conjunction with the luminosity densities that we derive
to estimate the stellar M/L ratio of the local Universe.  
Taking the stellar mass density of the Universe to
be $5.5 \pm 0.1 \pm 1.6 \times 10^8\,h\,M_{\sun}\,{\rm Mpc}^{-3}$, we find
$(3.64, 3.50, 3.05, 2.57, 2.00, 0.95)\pm 0.03M_{\sun}/L_{\sun}$ 
in $ugrizK$ passbands,
with systematic errors of $\sim 30$\%, owing to uncertainties
in SFHs and dust.
These values agree well with 0.93$M_{\sun}/L_{\sun}$
from \citet{cole01} using 2MASS $K$-band (although we account for 
systematic error in the stellar M/L ratios), 
and the range 2.6--5.2$M_{\sun}/L_{\sun}$
estimate of \citet{glazebrook03} in the $r$-band using the cosmic mean 
galaxy spectrum as a constraint.  Note that we scale both cases down
from a Salpeter IMF to our maximally-massive `diet' Salpeter IMF.
\citet{glazebrook03} find a slightly larger range of 
possible M/L ratios than we do.
Aperture effects may play a role, since they fit SFHs to 
the 3\arcsec fiber cosmic mean spectrum, which will be 
biased towards the inner (redder) parts of galaxies
\citep[e.g.,][]{bdj}.  Another systematic difference is that 
Glazebrook et al. construct the M/L ratio from the cosmic mean spectrum, which 
increases the uncertainties in how much light comes from old
stellar populations relative to the younger stars with lower M/L ratio.
We do not suffer as much from this uncertainty because
we see much of the old stellar population light directly
from red early-type galaxies.\footnote{Although note that \citet{rudnick03}
find that the sum of individual color-derived stellar masses for 
intensely star-forming galaxies can overpredict the total stellar mass
by nearly a factor of two, whereas estimating the stellar mass from the 
summed light of these star-forming galaxies yields a much more
accurate estimate, as the composite SF history of many bursty
galaxies is more nearly smooth than the SF histories of each 
individual galaxy.  This uncertainty will affect us only minimally, 
bearing in mind that the total stellar mass is dominated by 
early-type galaxies at the present epoch ($\ga $50\%), and that
the fraction of galaxies which are actively star-bursting is very 
low at the present epoch \citep[e.g.,][]{wolf03}.}

\section{Conclusions} \label{sec:conc}

A detailed understanding of the optical and NIR LFs and stellar 
MF is of fundamental importance to our understanding
galaxy formation and evolution.  We use a large sample
of galaxies from 2MASS and the SDSS EDR to estimate the 
optical and NIR LFs and stellar MF in the local Universe, assuming 
a universally-applicable stellar IMF.
We correct 2MASS extended source catalog
Kron magnitudes to total by comparing 2MASS magnitudes with deeper
data from a variety of sources.  Using the complete sky coverage of 2MASS 
to maximal effect, we find that the SDSS EDR
region is $\sim 8$\% overdense, and we correct for this overdensity 
when deriving LFs and stellar MFs.

We estimate $k$-corrections, evolution corrections, and
present-day stellar M/L ratios by comparison of galaxy 
$ugrizK$ magnitudes with galaxy evolution models 
at each galaxy's redshift.  The corrections and M/L ratios
we derive are in excellent agreement with previous work,
having the advantage that they incorporate the multi-passband
information to maximal effect.  We estimate $\sim$25\% systematic errors
in stellar M/L ratios from the effects of dust and bursts of SF, 
and we incorporate other random sources
of error in our analysis.  Assuming a 
universally-applicable stellar IMF, the limiting uncertainty
in stellar M/L ratios is the overall uncertainty in the number of 
faint, low-mass stars.  We conservatively adopt an IMF 
that has as many low-mass stars as possible without
violating dynamical constraints in nearby galaxies.
Of course, IMFs poorer
in low-mass stars are possible, and can be well-approximated
by subtracting a logarithmic constant from all stellar M/L ratios and stellar masses
presented in this paper.

We construct optical and NIR LFs for galaxies in the local 
Universe using the $V/V_{\rm max}$ formalism, which has the 
key advantage that it does not assume that galaxies in all environments
have the same LF.  The optical and NIR LFs that we 
estimate for this sample of galaxies agree to within the 
uncertainties with most recent literature 
optical and NIR LFs.  We argue that 2MASS misses faint, LSB
galaxies, leading to underestimates of up to 25\% in 
the $K$-band luminosity density.
The optical and NIR luminosity
densities in the local Universe look to be well-constrained
to within $\sim 20$\%, and matches qualitatively
a `cosmic SFH' that peaks at early times 
and continues to the present day at a reduced rate, 
with metallicities of roughly solar.  

We estimate the stellar MF using both $K$-band and $g$-band 
limited galaxy samples.  The MFs we derive using both 
samples are consistent, given the bias against
faint LSB galaxies in the 2MASS catalog.
The $g$-band derived stellar 
MF goes down to lower stellar masses than the
$K$-band MF.  Both MFs agree with 
the \citet{cole01} stellar MF, if the same IMFs are assumed.
The faint end slope of the stellar MF is steeper than $\alpha = -1.1$,
reflecting the relatively low stellar M/L ratios of low mass 
galaxies.\footnote{Catalogs of stellar, estimated \hi and 
H$_2$ gas masses \citep[following][]{bary} for $13 \le r \le 17.5$
SDSS EDR galaxies, along with electronic versions of the tables in 
this paper, are given at: 
\texttt {http://www.mpia-hd.mpg.de/homes/bell/data.html}}

Under the assumption of a universally-applicable stellar IMF,
the stellar mass density in the local Universe is 
$\Omega_* h = 2.0\pm0.6\times 10^{-3}$, accounting for all
sources of systematic and random error, {\it except for IMF 
uncertainty}.  To change to a different IMF, we would reduce this estimate
by a constant factor, as given in the text.
Our stellar mass density estimate is consistent with 
earlier estimates, with the advantage that the systematic
trends and errors in stellar M/L ratios are accounted for in
this work.  
We find `cosmic' stellar M/L ratios of 
$(3.64, 3.50, 3.05, 2.57, 2.00, 0.95)$, in solar units in $ugrizK$, with
$\sim 30$\% systematic errors, owing to the possible effects
of bursts of SF and dust. 

Finally, we examine type-dependence in the optical and NIR
LFs and the stellar MF.  In agreement with previous work, we
find that the characteristic luminosity or mass of early-type
galaxies is larger than for later types, and the faint end 
slope is steeper for later types than for earlier types.
These results are robust to systematic differences in 
galaxy typing, although the overall numbers of early and late-type
galaxies are somewhat dependent on the exact typing algorithm.
Accounting for typing uncertainties, we estimate
that at least half, and perhaps as much as 3/4, 
of the stellar mass in the Universe is in 
early-type galaxies.

\acknowledgements

We are grateful for helpful discussions and correspondence with Stefano
Andreon, Ivan Baldry, Michael
Blanton, Roc Cutri, David Hogg, Tom Jarrett, Peder Norberg, 
Hans-Walter Rix, Paolo Salucci, 
Steve Schneider, Rae Stiening, and Michael Strauss.
EFB\ was supported by the European Research Training
Network {\it Spectroscopic and Imaging Surveys for Cosmology}.
DHM and MDW acknowledge support 
by JPL/NASA through the 2MASS core science projects.
NK and MDW acknowledge support by NSF grants AST-9988146 \& AST-0205969 and
by the NASA ATP grant NAG5-12038.
This publication makes use of data products from the 
{\it Two Micron All Sky Survey}, which is a joint project of the 
University of Massachusetts and the Infrared Processing and 
Analysis Center/California Institute of Technology, funded by 
the National Aeronautics and Space Administration and 
the National Science Foundation.
This publication also makes use of the {\it Sloan Digital
Sky Survey} (SDSS).
Funding for the creation and distribution of the SDSS 
Archive has been provided by the Alfred P.\ Sloan Foundation, the 
Participating Institutions, the National Aeronautics and Space Administration,
the National Science Foundation, the US Department of Energy, 
the Japanese Monbukagakusho, and the Max Planck Society.  The SDSS
Web site is \texttt {http://www.sdss.org/}.  The SDSS Participating
Institutions are the University of Chicago, Fermilab, the Institute 
for Advanced Study, the Japan Participation Group, the Johns Hopkins
University, the Max Planck Institut f\"ur Astronomie, the Max
Planck Institut f\"ur Astrophysik, New Mexico State University, 
Princeton University, the United States Naval Observatory, and 
the University of Washington. 
This publication also made use of NASA's Astrophysics Data System 
Bibliographic Services.

\appendix
\section{Mass-to-light ratios and galaxy colors}

As an aid to workers in the field, we present the 
$K$ and $g$-band distributions of stellar M/L ratio,
and the color--M/L ratio relations in the different SDSS and 2MASS passbands.

\begin{figure*}[tb]
\center{\includegraphics[scale=0.9, angle=0]{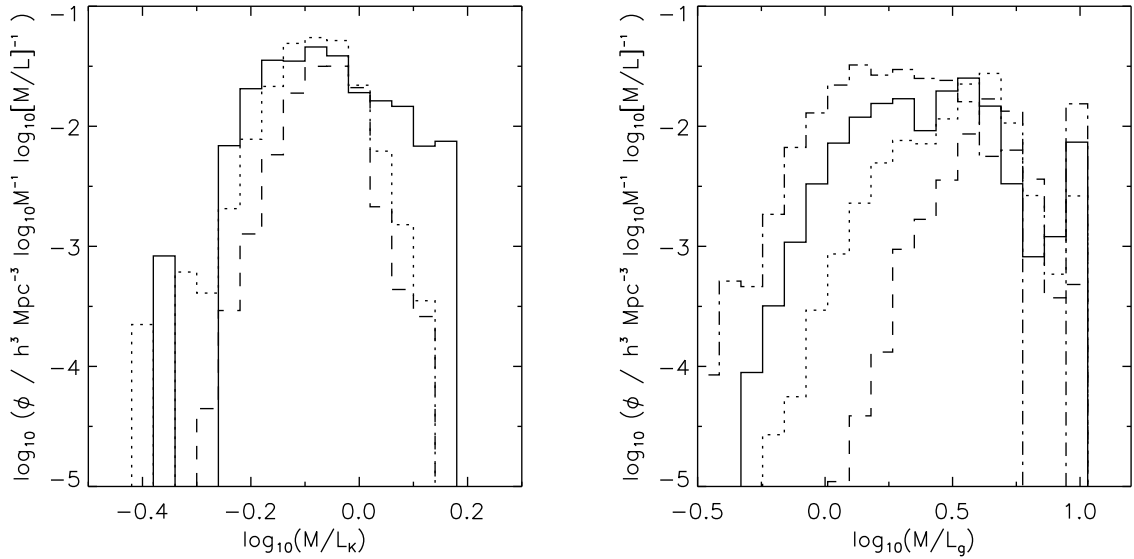}}
\caption{\label{fig:mldistrib} 
Distributions of stellar M/L ratio estimated from 
galaxy colors in $K$-band ({\it left}) and $g$-band ({\it right}).  
We show four
different galaxy stellar mass bins in units of solar mass ($M_{\sun}$):
$9 < \log_{10} Mh^2 \le 9.5$ ({\it dot-dashed}),
$9.5 < \log_{10} Mh^2 \le 10$ ({\it solid}),
$10 < \log_{10} Mh^2 \le 10.5$ ({\it dotted}),
and $10.5 < \log_{10} Mh^2 \le 11$ ({\it dashed}).  
The $K$-band $9 < \log_{10} Mh^2 \le 9.5$ bin is missing owing to poor
number statistics. }
\end{figure*}

\begin{figure}[tb]
\vspace{-0.5cm}
\hspace{-0.5cm}
\epsfbox{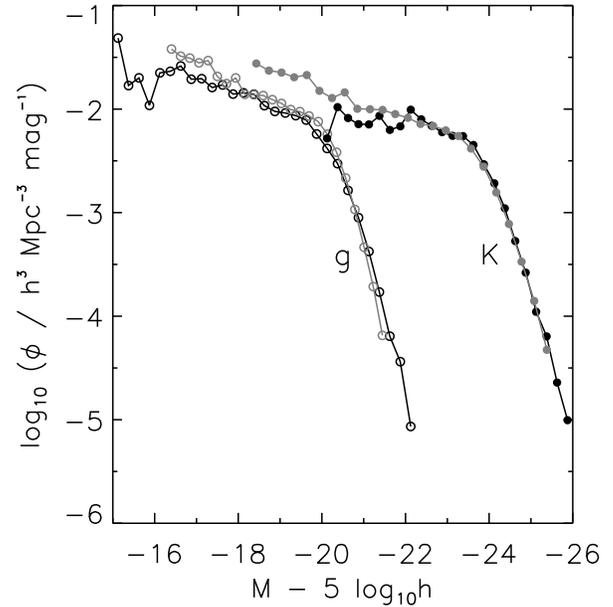}
\vspace{-0.2cm}
\caption{\label{fig:predlf} 
Comparison of the real $g$-band ({\it open circles}) and 
$K$-band ({\it solid circles}) LF, in black, with the 
predicted LFs transformed from the stellar MF using the average
stellar M/L ratio at a given stellar mass ({\it in grey}).   }
\end{figure}

\begin{figure*}[tb]
\center{\includegraphics[scale=0.8, angle=0]{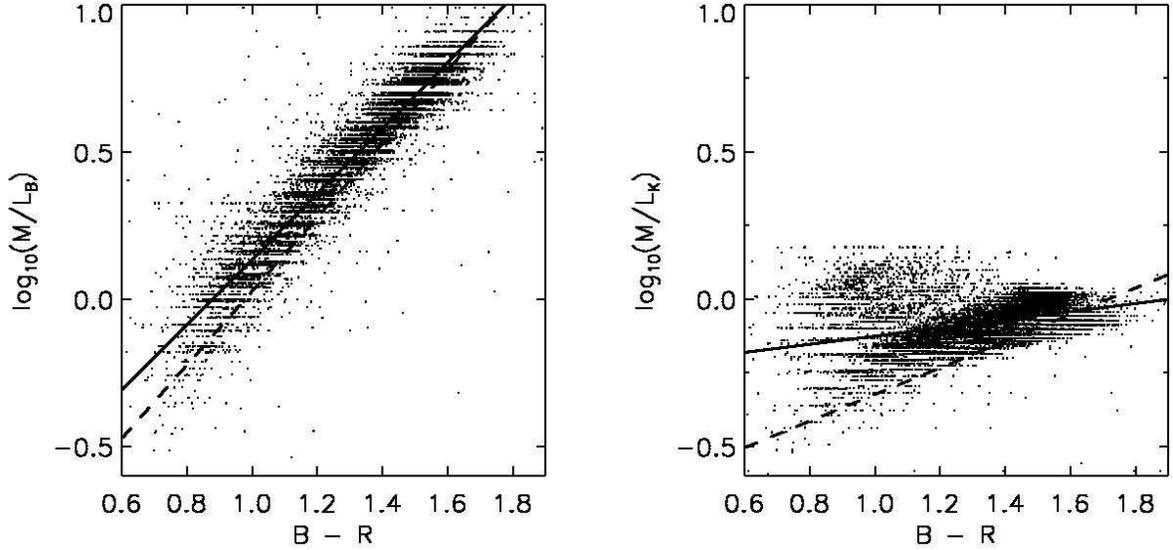}}
\caption{\label{fig:mlcomp} 
Comparison of estimated $B$-band and $K$-band stellar M/L ratios 
as a function of $B-R$ color 
for galaxies in this paper ({\it dots}).  In both panels we show 
a `robust' bi-square weighted
line fit ({\it solid line}), and the galaxy model color-M/L ratio correlations
({\it dashed line}) from \citet{ml}.   }
\end{figure*}

\begin{table}
\caption{Stellar M/L ratio distributions {\label{tab:ml}}}
\begin{center}
\begin{tabular}{lcccc}
\hline
\hline
M/L & \multicolumn{4}{c}{$\log_{10} (\phi/ h^3 {\rm Mpc^{-3}} \log_{10}M^{-1} \log_{10}[M/L]^{-1})$} \\
(1) & (2) & (3) & (4) & (5) \\
\hline
\multicolumn{5}{c}{K-band} \\
\hline
$-$0.40 & \nd & \nd & $-$3.65 & \nd\\
$-$0.36 & \nd & $-$3.08 & \nd & \nd\\
$-$0.32 & $-$2.16 & \nd & $-$3.21 & \nd\\
$-$0.28 & \nd & \nd & $-$3.39 & $-$4.35\\
$-$0.24 & $-$1.64 & $-$2.16 & $-$2.69 & $-$3.53\\
$-$0.20 & $-$1.97 & $-$1.69 & $-$2.11 & $-$2.90\\
$-$0.16 & $-$1.74 & $-$1.45 & $-$1.67 & $-$2.24\\
$-$0.12 & $-$1.83 & $-$1.46 & $-$1.31 & $-$1.72\\
$-$0.08 & $-$1.43 & $-$1.34 & $-$1.26 & $-$1.50\\
$-$0.04 & $-$1.86 & $-$1.41 & $-$1.29 & $-$1.50\\
 0.00 & $-$2.16 & $-$1.72 & $-$1.66 & $-$1.68\\
 0.04 & $-$1.66 & $-$1.79 & $-$2.21 & $-$2.67\\
 0.08 & $-$1.49 & $-$1.83 & $-$2.82 & $-$3.36\\
 0.12 & $-$1.50 & $-$2.17 & $-$3.45 & $-$3.58\\
 0.16 & \nd & $-$2.12 & \nd & \nd\\
\hline
\multicolumn{5}{c}{g-band} \\
\hline
$-$0.46 & $-$4.07 & \nd & \nd & \nd\\
$-$0.37 & $-$3.29 & \nd & \nd & \nd\\
$-$0.29 & $-$3.34 & $-$4.05 & \nd & \nd\\
$-$0.20 & $-$2.73 & $-$3.50 & $-$4.57 & \nd\\
$-$0.12 & $-$2.18 & $-$2.97 & $-$4.25 & \nd\\
$-$0.03 & $-$1.89 & $-$2.48 & $-$3.53 & \nd\\
 0.05 & $-$1.66 & $-$2.14 & $-$3.06 & $-$4.96\\
 0.14 & $-$1.49 & $-$1.92 & $-$2.64 & $-$4.41\\
 0.22 & $-$1.57 & $-$1.81 & $-$2.31 & $-$3.88\\
 0.31 & $-$1.53 & $-$1.77 & $-$2.12 & $-$3.02\\
 0.39 & $-$1.60 & $-$2.04 & $-$2.15 & $-$2.78\\
 0.48 & $-$1.62 & $-$1.71 & $-$1.94 & $-$2.45\\
 0.56 & $-$1.80 & $-$1.60 & $-$1.65 & $-$2.06\\
 0.65 & $-$2.25 & $-$1.83 & $-$1.56 & $-$1.77\\
 0.73 & $-$2.20 & $-$2.48 & $-$1.97 & $-$1.87\\
 0.82 & \nd & $-$3.09 & $-$2.58 & $-$2.44\\
 0.90 & \nd & $-$2.92 & $-$3.23 & $-$3.43\\
 0.99 & $-$1.81 & $-$2.13 & $-$2.58 & $-$3.32\\
\hline
\end{tabular}
\\ 
\vspace{-1.0cm}
\tablecomments{The distribution of stellar M/L ratio (1) in solar units is given \\
for different bins of mass in solar units: \\
(2) $9 < \log_{10} Mh^2 \le 9.5$;
(3) $9.5 < \log_{10} Mh^2 \le 10$;  \\
(4) $10 < \log_{10} Mh^2 \le 10.5$;
(5) $10.5 < \log_{10} Mh^2 \le 11$ }
\end{center}
\end{table}

\begin{table*}
\begin{footnotesize}
\begin{center}
\caption{Stellar M/L ratio as a function of color 
        {\label{tab:mltab}}}
\begin{tabular}{lcccccccccccccc}
\tableline
\tableline
{Color} & {$a_g$} & {$b_g$} 
& {$a_r$} & {$b_r$} 
& {$a_i$} & {$b_i$} 
& {$a_z$} & {$b_z$}  
& {$a_J$} & {$b_J$}  
& {$a_H$} & {$b_H$}  
& {$a_K$} & {$b_K$} \\
\tableline
$u - g$ & $-$0.221 &  0.485 & $-$0.099 &  0.345 & $-$0.053 &  0.268 & $-$0.105 &  0.226 & $-$0.128 &  0.169 & $-$0.209 &  0.133 & $-$0.260 &  0.123 \\
$u - r$ & $-$0.390 &  0.417 & $-$0.223 &  0.299 & $-$0.151 &  0.233 & $-$0.178 &  0.192 & $-$0.172 &  0.138 & $-$0.237 &  0.104 & $-$0.273 &  0.091 \\
$u - i$ & $-$0.375 &  0.359 & $-$0.212 &  0.257 & $-$0.144 &  0.201 & $-$0.171 &  0.165 & $-$0.169 &  0.119 & $-$0.233 &  0.090 & $-$0.267 &  0.077 \\
$u - z$ & $-$0.400 &  0.332 & $-$0.232 &  0.239 & $-$0.161 &  0.187 & $-$0.179 &  0.151 & $-$0.163 &  0.105 & $-$0.205 &  0.071 & $-$0.232 &  0.056 \\
$g - r$ & $-$0.499 &  1.519 & $-$0.306 &  1.097 & $-$0.222 &  0.864 & $-$0.223 &  0.689 & $-$0.172 &  0.444 & $-$0.189 &  0.266 & $-$0.209 &  0.197 \\
$g - i$ & $-$0.379 &  0.914 & $-$0.220 &  0.661 & $-$0.152 &  0.518 & $-$0.175 &  0.421 & $-$0.153 &  0.283 & $-$0.186 &  0.179 & $-$0.211 &  0.137 \\
$g - z$ & $-$0.367 &  0.698 & $-$0.215 &  0.508 & $-$0.153 &  0.402 & $-$0.171 &  0.322 & $-$0.097 &  0.175 & $-$0.117 &  0.083 & $-$0.138 &  0.047 \\
$r - i$ & $-$0.106 &  1.982 & $-$0.022 &  1.431 &  0.006 &  1.114 & $-$0.052 &  0.923 & $-$0.079 &  0.650 & $-$0.148 &  0.437 & $-$0.186 &  0.349 \\
$r - z$ & $-$0.124 &  1.067 & $-$0.041 &  0.780 & $-$0.018 &  0.623 & $-$0.041 &  0.463 & $-$0.011 &  0.224 & $-$0.059 &  0.076 & $-$0.092 &  0.019 \\
\tableline
{Color} & {$a_B$} & {$b_B$} 
& {$a_V$} & {$b_V$} 
& {$a_R$} & {$b_R$} 
& {$a_I$} & {$b_I$}  
& {$a_J$} & {$b_J$}  
& {$a_H$} & {$b_H$}  
& {$a_K$} & {$b_K$} \\
\tableline
$B - V$ & $-$0.942 &  1.737 & $-$0.628 &  1.305 & $-$0.520 &  1.094 & $-$0.399 &  0.824 & $-$0.261 &  0.433 & $-$0.209 &  0.210 & $-$0.206 &  0.135 \\
$B - R$ & $-$0.976 &  1.111 & $-$0.633 &  0.816 & $-$0.523 &  0.683 & $-$0.405 &  0.518 & $-$0.289 &  0.297 & $-$0.262 &  0.180 & $-$0.264 &  0.138 \\
\tableline \\
\end{tabular} 
\\ 
\vspace{-1.3cm}
\tablecomments{Stellar M/L ratios are given by
 $\log_{10}({\rm M/L}) = a_{\lambda} + (b_{\lambda} \times {\rm Color})$
where the M/L ratio is in solar units. 
If {\it all} galaxies are
sub-maximal then the above zero points ($a_{\lambda}$)
should be modified by subtracting an IMF dependent
constant as follows:
0.15 dex for a Kennicutt or Kroupa IMF, and 
0.4 dex for a Bottema IMF.  Scatter in the above correlations is
$\sim 0.1$ dex for all optical M/L ratios, and 0.1--0.2 dex for NIR 
M/L ratios (larger for galaxies with blue optical colors).
SDSS filters are in AB magnitudes; Johnson BVR and JHK are
in Vega magnitudes.}
\end{center}
\end{footnotesize}
\end{table*}

\subsection{The Distribution of Stellar M/L ratios}

In Fig.\ \ref{fig:mldistrib}, we show the number density of galaxies
as a function of their estimated stellar M/L ratio, assuming our 
maximal `diet' Salpeter IMF.  We denote different mass bins 
(in $M_{\sun}$) with 
different line styles: dot-dashed $9 < \log_{10} Mh^2 \le 9.5$,
solid $9.5 < \log_{10} Mh^2 \le 10$,
dotted $10 < \log_{10} Mh^2 \le 10.5$,
and dashed $10.5 < \log_{10} Mh^2 \le 11$.
In $K$-band, the $9 < \log_{10} Mh^2 \le 9.5$ bin is missing
because of poor number statistics.  In $g$-band, it is clear that 
the average stellar M/L ratio increases with increasing galaxy stellar mass, which
indicates that more of the stars were formed at an 
earlier time \citep[e.g.,][]{kauffmann03b}.  Moreover, the scatter becomes
somewhat narrower in stellar M/L ratio at high stellar
mass indicating less diversity in SFH.  
Massive galaxies tend to be rather old, regardless
of morphological type, whereas less massive galaxies can have a wide
range of ages \citep[see also, e.g., Fig.\ 2 of][]{kauffmann03b}.
Hints of this trend in $K$-band are visible, but much weaker, 
showing the well-documented lack of sensitivity of M/L$_K$ ratio to 
SFH \citep{ml}.
Sources of error include uncertainties from magnitude 
errors, systematic uncertainties in stellar M/L ratio from 
dust and bursts of SF ($\sim 25$\% in terms of stellar M/L ratio),
and Poisson errors.  The systematic uncertainties dominate, but are 
difficult to meaningfully estimate; thus, error bars are not given 
in this particular case.  We tabulate the $g$ and $K$-band distributions
in Table \ref{tab:ml}.

It is worth briefly commenting on why these distributions are 
useful.  In Fig.\ \ref{fig:predlf}, we show the observed
$g$-band ({\it open circles}) and $K$-band ({\it filled circles}) 
LFs in black.  Overplotted in grey are the predictions 
from the $g$-band-derived stellar MF, using the average 
stellar M/L ratio as a function of stellar mass (as is often 
used by galaxy modelers to transform a stellar mass distribution 
into a luminosity function\footnote{Often, a single `typical' 
stellar M/L ratio is used, which is even worse than the case we explore.}).  
To transform the stellar mass into 
luminosities, we adopt the bi-weight fit of 
stellar M/L ratio as a function of stellar mass: 
$\log_{10} ({\rm M/L}_g) = -2.61
+ 0.298 \log_{10} (M_*h^2/M_{\sun})$, and $\log_{10} ({\rm M/L}_K) = -0.42$\\
$+ 0.033 \log_{10} (M_*h^2/M_{\sun})$.  The estimated $K$-band LF is
in excellent agreement with the observed LF around the knee of the 
LF, and is in reasonable agreement at all luminosities with the
predicted $K$-band LF, once 2MASS's selection bias against
LSB galaxies is corrected for (the thick 
grey dashed line in Fig.\ \ref{fig:kbandtotal}).  This indicates
that the variation in stellar M/L ratio in $K$-band at a given stellar mass 
is sufficiently small so that the predicted LF is close to the real LF.
In contrast, using the average M/L$_g$ ratio at a given stellar mass
is clearly insufficient to reproduce
the $g$-band LF; especially at the faint end 
where one sees in Fig.\ \ref{fig:mldistrib} that the scatter
in M/L$_g$ ratio is particularly large.  This shows the importance of 
accounting for the spread in stellar M/L ratio at a given mass
when transforming a stellar mass function into a luminosity function, 
especially in the optical.\footnote{This also applies to transforming
LFs from one passband into another using a luminosity-dependent
typical color.  If the spread in color at a given luminosity is rather small,
one can get away with this approximation.  If the spread in color
is larger, the estimated LF will be biased.}

\subsection{The Stellar M/L Ratio--Color Correlation}

\citet{ml} presented relationships between Johnson/Cousins optical-NIR 
colors and stellar M/L ratios in the optical and NIR using the galaxy 
models of \citet{bell00}.  In this paper, we construct stellar 
M/L ratio estimates using galaxy evolution model fits to 
SDSS $ugriz$ and 2MASS $K$-band fluxes, finding
tight correlations between optical color and stellar M/L ratio (e.g.\
Fig.\ \ref{fig:ml}).  Therefore, in this appendix we 
compare our results with \citet{ml}, we examine the
color--M/L ratio correlations in detail where necessary, 
and we present fits to the 
color--M/L ratio correlations in the SDSS/2MASS passbands.

To facilitate inter-comparison between our results, 
we choose to predict Johnson/Cousins M/L ratio 
values for the best-fit SEDs to the SDSS/2MASS galaxies 
in the $g$-band selected sample.  We estimate $B-R$ colors using the 
transformation in \citet{fuku96}: $B - R = 1.506(g-r) + 0.370$,
with a $\sim 0.05$ mag systematic error.

We show two representative color--M/L ratio correlations
in Fig.\ \ref{fig:mlcomp}, where we show the 
M/L ratio in $B$-band ({\it left-hand panel}) 
and $K$-band ({\it right-hand panel}) as a function of $B - R$
color.  Additionally, we give the least-squares `robust' bi-weight fit to the 
estimates ({\it solid line}), and the relationship from 
the mass-dependent formation epoch with bursts model from 
\citet[{\it dashed line}]{ml}.  These two panels
are easily compared with panel {(\it d)} of Fig.\ 1 in \citet{ml}; indeed,
this is why we chose to estimate $B-R$ colors and Johnson/Cousins
M/L ratios.  The agreement between
the `least-squares fit to many passbands' methodology of this
paper is in excellent agreement with the galaxy modeling of 
\citet{ml} for the $B$-band M/L ratio--color relation.  This result 
is insensitive to detailed passband choice since all the optical M/L ratio--color
correlations that we derive are consistent with \citet{ml}.

In the right-hand panel of Fig.\ \ref{fig:mlcomp}, we show 
the run of $K$-band stellar M/L ratio estimates from the 
`least-squares fit to many passbands' methodology we adopt in this 
paper ({\it points}) against
the galaxy model-based estimate of \citet[{\it dashed line}]{ml}.  We note
the somewhat poorer agreement between these two different
methodologies.  There is a zero point offset, owing to our 
use of \pegase rather than the Bruzual \& Charlot (in preparation) 
models.   Furthermore, there is considerably more scatter 
at the blue end of the correlation than the models of \citet{ml}
predict, and a somewhat shallower correlation than is expected
on the basis of their galaxy modeling.  The data points fill
in the range of possible colors and M/L ratios of stellar populations with a wide
variety of ages and metallicities 
\citet[Fig.\ 2, panel {\it c}]{ml},
indicating that this spread is primarily caused by a
spread in metallicity .  In particular, it is clear that optically-blue galaxies
have a wide range of estimated metallicities.  Recall that 
our methodology in this paper is to estimate
ages and metallicities using the optical--NIR colors following \citet{bdj}.
In contrast, the galaxy evolution models of \citet{ml} do not have a 
large metallicity spread, and therefore, do not reproduce this feature.
This large metallicity spread flattens the optical color-M/L$_K$ ratio
correlation, adding $\sim 0.2$ dex scatter at the blue end.  It is important
to note that this 0.2 dex scatter is no more than a factor of two in excess
of the scatter in optical M/L ratios as a function of color.

In conclusion, we find that the optical M/L ratios as a function of 
color are in good agreement with \citet{ml}.  However, for
NIR M/L ratios we find that real galaxies suggest a larger metallicity
scatter than accounted for by \citet{ml}, leading to 
a shallower color-M/L ratio slope and a larger spread at the blue end.
In Table \ref{tab:mltab} we present
the correlations between SDSS $ugriz$ colors and SDSS/2MASS M/L ratios, 
and between $BVR$ colors and Johnson/Cousins M/L ratios,
to allow intercomparison with \citet{ml}, and to allow the
estimation of systematic differences between their work and ours
in the NIR.  Typical M/L ratio uncertainties are $\sim 0.1$ dex in 
the optical, and 0.1 (0.2) dex in the NIR at the red (blue) end.
We do not present $u$-band M/L estimates for $u$-band (because of its
strong sensitivity to recent SF) or correlations involving only
NIR colors (because of their strong metallicity sensitivity).


\begin{thebibliography}{}

\bibitem[Andreon(2002)]{andreon02}
	Andreon, S. 2002, \aap, 382, 495

\bibitem[Athanassoula, Bosma, \& Papaioannou(1987)]{ath87}
        Athanassoula, E., Bosma, A., \& Papaioannou, S. 1987, \aap, 179, 23

\bibitem[Baldry et al.(2002)]{baldry02}
        Baldry, I.\ K., et al. 2002, \apj, 569, 582

\bibitem[Bell(2003)]{radfir}
        Bell, E.\ F. 2003, \apj, 586, 794

\bibitem[Bell \& Bower(2000)]{bell00}
        Bell, E.\ F., \& Bower, R.\ G. 2000, \mnras, 319, 235 

\bibitem[Bell \& de Jong(2000)]{bdj}
        Bell, E.\ F., \& de Jong, R.\ S. 2000, \mnras, 312, 497 

\bibitem[Bell \& de Jong(2001)]{ml}
        Bell, E.\ F., \& de Jong, R.\ S. 2001, \apj, 550, 212 

\bibitem[Bell et al.(2003a)]{bary}
        Bell, E.\ F., McIntosh, D.\ H., Katz, N., \& Weinberg, M.\ D. 2003a, 
        \apj, 585, L117

\bibitem[Bell et al.(2003b)]{combo17}
	Bell, E.\ F., Wolf, C., Meisenheimer, K., Rix, H.-W., Borch, A.,
	Dye, S., Kleinheinrich, M., McIntosh, D.\ H. 2003, submitted to \apj
	{ }(astro-ph/0303394)

\bibitem[Benson et al.(2002)]{benson02}
        Benson, A.\ J., Lacey, C.\ G., Baugh, C.\ M., Cole, S., \& Frenk,
        C.\ S. 2002, \mnras, 333, 156

\bibitem[Bernardi et al.(2003a)]{bernardi_ii}
        Bernardi, M., et al. 2003a, \aj, 125, 1849

\bibitem[Bernardi et al.(2003b)]{bernardi_iii}
        Bernardi, M., et al. 2003b, \aj, 125, 1866

\bibitem[Bernardi et al.(2003c)]{bernardi_iv}
        Bernardi, M., et al. 2003c, \aj, 125, 1882

\bibitem[Blain et al.(1999)]{blain99}
        Blain, A.\ W., Smail, I., Ivison, R.\ J., \& Kneib, J.-P. 1999, 
        \mnras, 302, 632

\bibitem[Blanton et al.(2001)]{blanton01}
        Blanton, M.\ R., et al. 2001, \apj, 121, 2358

\bibitem[Blanton et al.(2003a)]{blanton03k}
        Blanton, M.\ R., et al. 2003a, \aj, 125, 2348

\bibitem[Blanton et al.(2003b)]{blanton03c}
        Blanton, M.\ R., et al. 2003b, submitted to \apj { }(astro-ph/0209479)

\bibitem[Blanton et al.(2003c)]{blanton03}
        Blanton, M.\ R., et al. 2003c, submitted to \apj { }(astro-ph/0210215)

\bibitem[Bottema(1993)]{bottema93}
        Bottema, R. 1993, \aap, 275, 16

\bibitem[Bottema(1997)]{bottema97}
        Bottema, R. 1997, \aap, 328, 517

\bibitem[Bottema(1999)]{bottema99}
        Bottema, R. 1999, \aap, 348, 77

\bibitem[Brinchmann \& Ellis(2000)]{brinchmann00}
        Brinchmann, J., \& Ellis, R.\ S. 2000, \apj, 536, 77L

\bibitem[Bromley et al.(1998)]{bromley98}
        Bromley, B.\ C., Press, W.\ H., Lin, H., Kirshner, R.\ P. 1998, 
        \apj, 505, 25

\bibitem[Boselli et al.(2000)]{bos00}
Boselli, A., Gavazzi, G., Franzetti, P., Pierini, D., \& Scodeggio, M. 2000,
  \aap, 142, 73

\bibitem[Bower, Lucey, \& Ellis(1992)]{ble92}
        Bower, R.\ G., Lucey, J.\ R., \& Ellis, R.\ S. 1992, \mnras, 254, 601

\bibitem[Cole et al.(2000)]{cole00}
        Cole, S., Lacey, C., Baugh, C.\ M., Frenk, C.\ S. 2000, 
        \mnras, 319, 168

\bibitem[Cole \& Lacey(1996)]{cole96}
        Cole, S., Lacey, C.\ G. 1996, \mnras, 281, 716

\bibitem[Cole et al.(2001)]{cole01}
        Cole, S., et al. 2001, \mnras, 326, 255

\bibitem[Colless et al.(2001)]{colless01}
        Colless, M., et al. 2001, \mnras, 328, 1039

\bibitem[Courteau \& Rix(1999)]{courteau99}
        Courteau, S., \& Rix, H.-W. 1999, \apj, 513, 561

\bibitem[Cretton, Rix, \& de Zeeuw(2000)]{cretton00}
        Cretton, N., Rix, H.-W., \& de Zeeuw, P.\ T. 2000, \apj, 536, 319

\bibitem[Cross \& Driver(2002)]{cross02}
        Cross, N., \& Driver, S.\ P. 2002, \mnras, 329, 579

\bibitem[de Blok \& McGaugh(1998)]{deblok98}
        de Blok, W.\ J.\ G., \& McGaugh, S.\ S. 1998, \apj, 508, 132

\bibitem[de Jong \& Lacey(2000)]{dejong00}
        de Jong, R.\ S., \& Lacey, C. 2000, \apj, 545, 781

\bibitem[De Propris et al.(2003)]{deprop03}
        De Propris, R., et al. 2003, \mnras, 342, 725

\bibitem[Efstathiou et al.(1988)]{eep}
        Efstathiou, G., Ellis, R.\ S., \& Peterson, B.\ A. 1988, \mnras, 
        232, 431

\bibitem[Felten(1977)]{felten}
        Felten, J.\ E. 1977, \aj, 82, 861

\bibitem[Freedman et al.(2001)]{freedman01}
        Freedman, W.\ L., et al. 2001, \apj, 553, 47

\bibitem[Fukugita, Hogan, \& Peebles(1998)]{fuku98}
        Fukugita, M., Hogan, C.\ J., \& Peebles, P.\ J.\ E. 1998, 
        \apj, 503, 518

\bibitem[Fukugita et al.(1996)]{fuku96}
        Fukugita, M., Ichikawa, T., Gunn, J.\ E., Doi, M., Shimasaku, K., 
        \& Schneider, D.\ P. 1996, \aj, 111, 1748 

\bibitem[Fioc \& Rocca-Volmerange(1997)]{fioc97}
        Fioc, M., \& Rocca-Volmerange, B. 1997, \aap, 326, 950

\bibitem[Gardner et al.(1997)]{gardner97}
        Gardner, J.\ P., Sharples, R.\ M., Frenk, C.\ S., \& Carrasco, B.\ E.
        1997, 480L, 99

\bibitem[Gavazzi et al.(1996a)]{gav96a}
        Gavazzi, G., Pierini, D., Boselli, A., \& Tuffs, R. 1996a, 
        \aap, 120, 489

\bibitem[Gavazzi et al.(1996b)]{gav96b}
Gavazzi, G., Pierini, D., Baffa, C., Lisi, F., Hunt, L.\ K., Randone, I.,
  \& Boselli, A. 1996b, \aap, 120, 521

\bibitem[Gavazzi et al.(2000)]{gav00}
Gavazzi, G., Franzetti, P., Scodeggio, M., Boselli, A., Pierini, D., Baffa, C.,
  Lisi, F., \& Hunt, L.\ K. 2000, \aap, 142, 65

\bibitem[Glazebrook et al.(1995)]{glazebrook95}
        Glazebrook, K., Peacock, J.\ A., Miller, L., \& Collins, C.\ A.
        1995, \mnras, 275, 169

\bibitem[Glazebrook et al.(2003)]{glazebrook03}
        Glazebrook, K., et al. 2003, \apj, 587, 55

\bibitem[Gordon et al.(1997)]{gordon97}
        Gordon, K.\ D., Calzetti, D., Witt, A.\ N. 1997, \apj, 487, 625

\bibitem[Gould, Bahcall, \& Flynn(1997)]{gould97}
        Gould, A., Bahcall, J.\ N., \& Flynn, C. 1997, \apj, 482, 913

\bibitem[Haarsma et al.(2000)]{haarsma00}
        Haarsma, D.\ B., Partridge, R.\ B., Windhorst, R.\ A., Richards, 
        E.\ A. 2000, \apj, 544, 641

\bibitem[Hogg et al.(2002)]{hogg02}
        Hogg, D.\ W., et al. 2002, \aj, 124, 646

\bibitem[Hogg et al.(2003)]{hogg03}
        Hogg, D.\ W., et al. 2003, \apj, 585, L5

\bibitem[Huang et al.(2003)]{huang03}
        Huang, J.-S., Glazebrook, K., Cowie, L.\ L., \& Tinney, C.
        2003, \apj, 584, 203

\bibitem[H\"utsi et al.(2003)]{hutsi03}
        H\"utsi, G., Einasto, J., Tucker, D.\ L., Saar, E., Einasto, M., 
        M\"uller, V., Hein\"am\"aki, P., \& Allam, S.\ S. 2003, submitted
        to \aap{ }(astro-ph/0212327)

\bibitem[Jarrett et al.(2000)]{jarrett00}
        Jarrett, T. H., Chester, T., Cutri, R., Schneider, S., Skrutskie, M., 
        \& Huchra, J. P. 2000, \aj, 119, 2498

\bibitem[Kauffmann et al.(2003a)]{kauffmann03a}
        Kauffmann, G., et al. 2003a, \mnras, 341, 33

\bibitem[Kauffmann et al.(2003b)]{kauffmann03b}
        Kauffmann, G., et al. 2003b, \mnras, 341, 54

\bibitem[Kennicutt(1983)]{kennicutt83} 
        Kennicutt Jr., R. C. 1983, \apj, 272, 54

\bibitem[Kochanek et al.(2001)]{kochanek01}
        Kochanek, C.\ S., et al. 2001, \apj, 560, 566

\bibitem[Kodama \& Arimoto(1997)]{ka97}
        Kodama, T., \& Arimoto, N. 1997, \aap, 320, 41

\bibitem[Kranz, Slyz, \& Rix(2003)]{kranz03}
        Kranz, T., Slyz, A., \& Rix, H.-W. 2003, \apj, 586,143 

\bibitem[Kron(1980)]{kron}
        Kron, R.\ G. 1980, \apjs, 43, 305

\bibitem[Kroupa, Tout \& Gilmore(1993)]{kroupa93}
        Kroupa, P., Tout, C.\ A., \& Gilmore, G. 1993, \mnras, 262, 545

\bibitem[Kroupa(2002)]{kroupa02}
        Kroupa, P. 2002, Science, 295, 82

\bibitem[Lee et al.(2003)]{lee03}
	Lee, H., McCall, M.\ L., Kingsburgh, R.\ L., Ross, R., \& Stevenson,
	C.\ C. 2003, \aj, 125, 146

\bibitem[Lee, McCall \& Richer(2003)]{lee03_2}
	Lee, H., McCall, M.\ L., \& Richer, M.\ G. 
	2003, \aj, 125, 2975

\bibitem[Lemson \& Kauffmann(1999)]{lemson99}
        Lemson, G., Kauffmann, G. 1999, \mnras, 302, 111

\bibitem[Lilly et al.(1995)]{lilly95}
        Lilly, S.\ J., Tresse, L., Hammer, F., Crampton, D., Le Fevre, 
        O. 1995, \apj, 455, 108

\bibitem[Lin et al.(1996)]{lin96}
        Lin, H., Kirshner, R.\ P., Shectman, S.\ A., Landy, S.\ D., 
        Oemler, A., Tucker, D.\ L., \& Schechter, P.\ L. 1996, \apj, 464, 60

\bibitem[Lin et al.(1999)]{lin99}
        Lin, H., Yee, H.\ K.\ C., Carlberg, R.\ G., Morris, S.\ L., 
        Sawicki, M., Patton, D.\ R., Wirth, G., \& Shepherd, C.\ W. 1999, 
        \apj, 518, 533

\bibitem[Liske et al.(2003)]{liske03}
        Liske, J., Lemon, D.\ J., Driver, S.\ P., Cross, N.\ J.\ G., 
        \& Couch, W.\ J. 2002, submitted to \mnras { }(astro-ph/0207555)

\bibitem[Loveday(2000)]{loveday}
        Loveday, J. 2000, \mnras, 312, 557

\bibitem[Madau et al.(1996)]{madau96}
        Madau, P., Ferguson, H.\ C., Dickinson, M.\ E., 
        Giavalisco, M., Steidel, C.\ C., \& Fruchter, A. 1996, 
        \mnras, 283, 1388

\bibitem[Martin \& Kennicutt(2001)]{martin01} 
  Martin, C.\ L., Kennicutt Jr., R.\ C. 2001, \apj, 555, 301

\bibitem[Mo et al.(1998)]{mo98}
        Mo, H.\ J., Mao, S., \& White, S.\ D.\ M. 1998, \mnras, 295, 319

\bibitem[Nakamura et al.(2003)]{nakamura03}
        Nakamura, O., Fukugita, M., Yasuda, N., Loveday, J., Brinkmann,
        J., Schneider, D.\ P., Shimasaku, K., \& SubbaRao, M. 2003, \aj,  
        125, 1682

\bibitem[Nikolaev et al.(2000)]{nikolaev00}
        Nikolaev, S., Weinberg, M.\ D., Skrutskie, M.\ F., 
        Cutri, R.\ M., Wheelock, S.\ L., Gizis, J.\ E., \& Howard, E.\ M.
        2000, \aj, 120, 3340

\bibitem[Norberg et al.(2002)]{norberg02}
        Norberg, P., et al. 2002, \mnras, 332, 827

\bibitem[Persic \& Salucci(1992)]{persic92}
        Persic, M., \& Salucci, P. 1992, \mnras, 258, P14

\bibitem[Pier et al.(2003)]{pier03}
        Pier, J.\ R., et al. 2003, \aj, 125, 1559

\bibitem[Rudnick et al.(2003)]{rudnick03}
	Rudnick, G., et al. 2003, submitted to \apj{ }(astro-ph/0307149)

\bibitem[Salpeter(1955)]{salp}
        Salpeter, E.\ E. 1955, \apj, 121, 161

\bibitem[Salucci \& Persic(1999)]{salucci}
        Salucci, P., \& Persic, M. 1999, \mnras, 309, 923

\bibitem[Sandage et al.(1979)]{sty}
        Sandage, A., Tammann, G.\ A., \& Yahil, A. 1979, \apj, 232, 352

\bibitem[Scalo(1986)]{scalo86}
        Scalo, J. M. 1986, Fundam. Cosmic Phys., 11, 1

\bibitem[Scalo(1998)]{scalo98}
        Scalo, J. 1998, in `The Stellar Initial Mass Function 
        (38th Herstmonceux Conference)', eds. G.\ Gilmore and D.\ Howell,
        ASP Conference Series, v. 142, p 201.  

\bibitem[Schechter(1976)]{schechter}
        Schechter, P. 1976, \apj, 203, 297

\bibitem[Schlegel, Finkbeiner, \& Davis(1998)]{sfd}
        Schlegel, D.\ J., Finkbeiner, D.\ P., \& Davis, M. 1998, \apj,
        500, 525

\bibitem[Schweizer \& Seitzer(1992)]{schweizer92}
        Schweizer, F., and Seitzer, P. \ 1992, \aj, 104, 1039

\bibitem[Skrutskie et al.(1997)]{skrut}
        Skrutskie, M.\ F., et al. 1997, in `The Impact of 
        Large Scale Near-IR Sky Surveys', eds. F. Garzon et al., 
        p 25. (Dordrecht: Kluwer Academic Publishing Company)

\bibitem[Shimasaku et al.(2001)]{shim}
        Shimasaku, K., et al. 2001, \aj, 122, 1238

\bibitem[Somerville, Primack, \& Faber(2001)]{somerville01}
        Somerville, R.\ S., Primack, J.\ R., Faber, S.\ M. 2001, 
        \mnras, 320, 504

\bibitem[Stoughton et al.(2002)]{edr}
        Stoughton, C., et al. 2002, \aj, 123, 485

\bibitem[Strateva et al.(2001)]{strat}
        Strateva, I., et al. 2001, \aj, 122, 1861

\bibitem[Strauss et al.(2002)]{strauss02}
        Strauss, M.\ A., et al. 2002, \aj, 124, 1810

\bibitem[Trentham \& Tully(2002)]{trentham02}
        Trentham, N., Tully, R.\ B. 2002, \mnras, 335, 712

\bibitem[Tully \& Fisher(1977)]{tully77}
        Tully, R.\ B., \& Fisher, J.\ R. 1977, \aap, 54 661

\bibitem[Tully et al.(1998)]{tully98}
        Tully, R.\ B., Pierce, M.\ J., Huang, J.-S., Saunders, W.,
        Verheijen, M.\ A.\ W., \& Witchalls, P.\ L. 1998, \aj, 115, 2264

\bibitem[Tully et al.(2002)]{tully02}
        Tully, R.\ B., Somerville, R.\ S., Trentham, N., \&
        Verheijen, M.\ A.\ W. 2002, \apj, 569, 573

\bibitem[Weiner, Sellwood, \& Williams(2001)]{weiner01}
        Weiner, B.\ J., Sellwood, J.\ A., \& Williams, T.\ B. 2001, 
        \apj, 546, 931

\bibitem[Wolf et al.(2003)]{wolf03}
        Wolf, C., Meisenheimer, K., Rix, H.-W., Borch, A., Dye, S., \&
        Kleinheinrich, M. 2003, \aap, 401, 73

\bibitem[Wright(2001)]{wright01}
        Wright, E.\ L. 2001, \apj, 556, 17L

\bibitem[Yan et al.(1999)]{yan99}
        Yan, L., McCarthy, P.\ J., Freudling, W., Teplitz, H.\ I., 
        Malumuth, E.\ M., Weymann, R.\ J., Malkan, M.\ A. 1999, 
        \apj, 519, L47

\bibitem[York et al.(2000)]{york00}
  York, D.\ G., et al. 2000, \aj, 120, 1579

\bibitem[Zabludoff \& Mulchaey(2000)]{zabludoff00}
        Zabludoff, A.\ I., \& Mulchaey, J.\ S. 2000, \apj, 539, 136

\end{thebibliography}
\end{document}